\documentclass[oneside,american,index=totoc]{scrbook}
\usepackage[T1]{fontenc}
\usepackage[latin9]{inputenc}
\usepackage{babel}
\usepackage{float}
\usepackage{amsthm}
\usepackage{amsmath}
\usepackage{makeidx}
\makeindex
\usepackage{graphicx}
\usepackage[numbers]{natbib}
\usepackage[unicode=true,
 bookmarks=true,bookmarksnumbered=false,bookmarksopen=false,
 breaklinks=false,pdfborder={0 0 0},backref=false,colorlinks=false]
 {hyperref}
\hypersetup{
 pdfauthor={Alexander Streltsov}}

\makeatletter
\newcommand{\lyxaddress}[1]{
\par {\raggedright #1
\vspace{1.4em}
\noindent\par}
}
  \theoremstyle{definition}
  \newtheorem*{example*}{\protect\examplename}

\@ifundefined{date}{}{\date{}}
\usepackage{times}
\usepackage{txfonts}
\usepackage{braket}
\usepackage{pstricks}
\usepackage{dsfont}
\def \openone { \mathds{1} }

\makeatother

  \providecommand{\examplename}{Example}

\begin{document}

\title{Quantum Discord}

\subtitle{and its role in quantum information theory}

\author{Alexander Streltsov}

\maketitle

\lyxaddress{Alexander Streltsov\\
ICFO\textendash{}The Institute of Photonic Sciences\\
08860 Castelldefels (Barcelona), Spain\\
\href{mailto:alexander.streltsov@icfo.es}{alexander.streltsov@icfo.es}}

\vfill{}

The final publication is available at Springer via \href{http://dx.doi.org/10.1007/978-3-319-09656-8}{http://dx.doi.org/10.1007/978-3-319-09656-8}.

\chapter*{Acknowledgements}

Parts of this work are based on the research carried out during my
PhD at the Heinrich-Heine-Universität Düsseldorf \citep{Streltsov2013b}.
I thank Dagmar Bruß for the excellent supervision of the PhD, and
I also thank the following people for discussion and support: Gerardo
Adesso, Remigiusz Augusiak, Otfried Gühne, Hermann Kampermann, Matthias
Kleinmann, Maciej Lewenstein, Tobias Moroder, Marco Piani, and Wojciech
Zurek. I also acknowledge financial support by the Alexander von Humboldt-Foundation,
the John Templeton Foundation, EU IP SIQS, ERC AdG OSYRIS, and EU-Spanish
Ministry CHISTERA DIQIP.

\tableofcontents{}\printindex{}

\chapter*{Abstract}

Quantum entanglement is the most popular kind of quantum correlations,
and its fundamental role in several tasks in quantum information theory
like quantum cryptography, quantum dense coding, and quantum teleportation
is undeniable. However, recent results suggest that various applications
in quantum information theory do not require entanglement, and that
their performance can be captured by a new type of quantum correlations
which goes beyond entanglement. Quantum discord, introduced by Zurek
more than a decade ago, is the most popular candidate for such general
quantum correlations. In this work we give an introduction to this
modern research direction. After a short review of the main concepts
of quantum theory and entanglement, we present quantum discord and
general quantum correlations, and discuss three applications which
are based on this new type of correlations: remote state preparation,
entanglement distribution, and transmission of correlations. We also
give an outlook to other research in this direction.

\chapter{Introduction}

Quantum entanglement has fascinated the minds of physicists since
the very inception of quantum theory \citep{Schroedinger1935}. Entangled
quantum systems can behave in a bizarre way, exhibiting features which
seem to contradict ``our common sense notions of how the world works''
\citep[p. 114]{Nielsen2000}. This was first pointed out in a seminal
work by Einstein, Podolsky, and Rosen, who concluded that the quantum
theory must be incomplete \citep{Einstein1935}. However, about 30
years after Einstein's objection, Bell proposed an experiment, which
aimed to distinguish between predictions made by quantum theory on
the one hand, and Einstein's arguments on the other hand \citep{Bell1964}.
Bell's ideas served as a starting point for Clauser, Horne, Shimony,
and Holt, who formulated an inequality which is known today as the
CHSH inequality \citep{Clauser1969}. Following Einstein \emph{et
al.}, Nature should respect the CHSH inequality, and the fact that
it can be violated in quantum theory demonstrates the incompleteness
of quantum mechanics. 

Due to its simplicity, the CHSH inequality could be tested experimentally
by Freedman and Clauser already short time after its discovery \citep{Freedman1972}.
The data showed a violation of the CHSH inequality, thus invalidating
Einstein's arguments, in favor of the quantum mechanical description
of Nature. Later in the years 1981/82 Aspect \emph{et al.} performed
three experiments \citep{Aspect1981,Aspect1982,Aspect1982a}, confirming
the results of Freedman and Clauser. Since that time, several experiments
have demonstrated violation of the CHSH inequality, although some
loopholes still remained open \citep{Horodecki2009}.

The formal definition of entanglement as we use it today can be dated
back to the year 1989, when Werner extended the concept of entanglement
to all mixed quantum states \citep{Werner1989}. Werner's work can
be regarded as the starting point for the theory of entanglement,
which studies properties and implications of entanglement, and its
role in such fundamental tasks like quantum cryptography \citep{Ekert1991},
quantum dense coding \citep{Bennett1992}, and quantum teleportation
\citep{Bennett1993}. Several important contributions to the theory
of entanglement also came from the Horodecki family: one example is
the discovery of bound entanglement \citep{Horodecki1998}. Bound
entangled states need some amount of entanglement to be created, but
cannot be used for the extraction of any pure entangled state. A comprehensive
review on this topic can be found in \citep{Horodecki2009}.

The role of entanglement in quantum algorithms is still subject of
extensive debate. This is due to the results by Jozsa and Linden,
who showed that a quantum computer operating on a pure state needs
entanglement in order to have an exponential speedup compared to classical
computation \citep{Jozsa1997,Jozsa2003}. Although exponential speedup
of a quantum computer is not yet rigorously proven, there is strong
evidence for its existence. One of the most prominent examples pointing
in this direction is Shor's prime factorization algorithm proposed
in \citep{Shor1994}. The algorithm is able to find the prime factors
for any product of two primes on a quantum computer, where the time
for the computation grows polynomially in the number of the input
bits. This is significantly faster, compared to the best known classical
algorithm, which exhibits an exponential increase of the running time.

Due to the presence of entanglement in Shor's algorithm \citep{Jozsa2003}
one might be tempted to see entanglement as the key resource for quantum
computation. While for \emph{pure state} quantum computation this
is indeed the case, the situation becomes more involved if \emph{mixed
state} quantum computation is considered \citep{Jozsa2003}. A popular
example for mixed state quantum computation has been presented by
Knill and Laflamme \citep{Knill1998}. Surprisingly, their algorithm
is able to solve certain problems efficiently for which no efficient
classical algorithm is known even with vanishingly little entanglement
\citep{Datta2005}. This finding triggered the search for quantum
correlations beyond entanglement, which should be responsible for
the efficiency of a quantum computer.

\emph{Quantum discord}, introduced by Zurek in the year 2000, has
been recognized as a possible candidate for those general quantum
correlations \citep{Zurek2000,Ollivier2001}. On the one hand, quantum
discord can even exist in systems which are not entangled. On the
other hand, it has been shown that the algorithm presented by Knill
and Laflamme exhibits nonvanishing amount of discord \citep{Datta2008a}.
An even stronger statement has been made by Eastin, who showed that
mixed state quantum computation with zero discord in each step can
be simulated efficiently on a classical computer \citep{Eastin2010}.
Three years after Zurek has proposed quantum discord as a new kind
of quantum correlations beyond entanglement, he gave it an alternative
thermodynamical interpretation \citep{Zurek2003}. He considered the
amount of work which can be extracted from a quantum system by a classical
and a quantum Maxwell's demon. He showed that the quantum demon is
more powerful, since it can operate on the whole quantum state, while
the classical demon is restricted to local subsystems only. Zurek
concluded that more work can be extracted in the quantum case, and
this quantum advantage is related to the quantum discord.

Approximately at the same time when Zurek defined quantum discord,
a closely related quantity has been proposed by Henderson and Vedral
\citep{Henderson2001}. The authors aimed to separate correlations
into quantum and purely classical parts by postulating several reasonable
properties. This approach is significantly different from Zurek's,
and the fact that both arrive at the same result is surprising. Another
related quantity is the \emph{information deficit}, presented in \citep{Oppenheim2002}.
The authors study the amount of work, which can be extracted from
a heat bath using a mixed quantum state. If the mixed state is shared
by two parties, the amount of extractable work is usually smaller,
compared to the case where the whole state is in possession of a single
party. The difference of these two quantities is the information deficit.

In the light of these results, it is not surprising that in the last
few years an enormous amount of research has been devoted to tasks
in quantum information theory which are not based on entanglement
\citep{Modi2012}. Quantum discord has been related to the performance
of some of those tasks: remote state preparation \citep{Dakic2012}
and information encoding \citep{Gu2012} being popular examples. Experimental
techniques for detecting general quantum correlations have also been
presented \citep{Gessner2014}. In this work, we give an introduction
to general quantum correlations beyond entanglement and present a
detailed discussion on their role for remote state preparation \citep{Dakic2012},
entanglement distribution \citep{Streltsov2012a,Chuan2012}, and transmission
of correlations \citep{Streltsov2013a,Brandao2013}. We start by briefly
reviewing the mathematical framework of quantum theory, followed by
a short introduction to quantum entanglement. After introducing quantum
discord and related quantifiers of quantum correlations, we discuss
their role in quantum information theory, and also present a short
outlook on other research directions.

\chapter{Quantum theory}

\section{Quantum states}

In quantum mechanics, any physical system is completely described
by a state vector $\ket{\Psi}$ in a Hilbert space ${\cal H}$. A
system with a two-dimensional Hilbert space is also called a \emph{qubit\index{Qubit}}
(quantum bit). If not otherwise stated, we consider a Hilbert space
with an arbitrary but finite dimension. For two parties, Alice ($A$)
and Bob ($B$), with Hilbert spaces ${\cal H}_{A}$ and ${\cal H}_{B}$
the total Hilbert space is a tensor product of the subsystem spaces:
${\cal H}_{AB}={\cal H}_{A}\otimes{\cal H}_{B}$. 

Any system which is described by a single state vector is said to
be in a \emph{pure state\index{Pure state}}. However, in a realistic
experimental setup the physical state of the considered system is
not completely known. If the system is in the pure state $\ket{\psi_{i}}$
with probability $p_{i}$, the physical state of the system can be
described using the \emph{density operator\index{Density operator}}
\begin{equation}
\rho=\sum_{i}p_{i}\ket{\psi_{i}}\bra{\psi_{i}}.
\end{equation}
The state of such a system is called \emph{mixed state\index{Mixed state}}.
In the following, whenever we talk about quantum states, we usually
mean mixed states. 

In order to have a meaningful physical interpretation, any density
operator has the following two properties: 
\begin{itemize}
\item $\rho$ has trace equal to one: 
\begin{equation}
\mathrm{Tr}[\rho]=1,\label{eq:trace}
\end{equation}

\item $\rho$ is a positive operator: 
\begin{equation}
\braket{\psi|\rho|\psi}\geq0\label{eq:positivity}
\end{equation}
 for any vector $\ket{\psi}$.
\end{itemize}
Note that the second property also implies that $\rho$ is Hermitian:
$\rho^{\dagger}=\rho$. These two condition are essential for the
definition of quantum measurements and operations, which is presented
in the following.

\section{Quantum measurements and operations}

\index{Quantum!measurement}Quantum measurement is one of the most
important concepts in quantum theory. Most physicists are familiar
with the \index{Projective measurement}\emph{projective measurement}:
for a spin-$\frac{1}{2}$ particle in the state 
\begin{equation}
\ket{\psi}=a\ket{\uparrow}+b\ket{\downarrow},\label{eq:spin1/2}
\end{equation}
the probability to measure ``spin up'' or ``spin down'' is given
by $p(\uparrow)=\left|a\right|^{2}$ or $p(\downarrow)=\left|b\right|^{2}=1-p(\uparrow)$.
Moreover, the measurement postulate of quantum mechanics tells us
that the quantum state after the measurement is either $\ket{\uparrow}$
or $\ket{\downarrow}$, depending on the outcome of the measurement.

In quantum information theory, a more general definition is considered.
A general quantum measurement is described by a collection $\left\{ E_{i}\right\} $
of \emph{measurement operators}\index{Measurement operator} that
satisfy the completeness equation: 
\begin{equation}
\sum_{i}E_{i}^{\dagger}E_{i}=\openone,\label{eq:completeness}
\end{equation}
where $\openone$ is the identity operator. Given a density operator
$\rho$ and the set of measurement operators $\left\{ E_{i}\right\} $,
the probability that the result $i$ occurs is given by 
\begin{equation}
p_{i}=\mathrm{Tr}[E_{i}^{\dagger}E_{i}\rho].\label{eq:pi}
\end{equation}
After the measurement with outcome $i$, the state of the system is
described by the density operator 
\begin{equation}
\rho_{i}=\frac{1}{p_{i}}(E_{i}\rho E_{i}^{\dagger}).
\end{equation}
The set of operators 
\begin{equation}
M_{i}=E_{i}^{\dagger}E_{i}
\end{equation}
is also called positive operator-valued measure (\emph{POVM})\index{POVM}.
Due to the completeness equation (\ref{eq:completeness}), the POVM
elements $M_{i}$ sum up to the identity operator: $\sum_{i}M_{i}=\openone$.
Moreover, due to Eq. (\ref{eq:pi}) the probabilities $p_{i}$ can
also be obtained from the POVM elements $M_{i}$: $p_{i}=\mathrm{Tr}[M_{i}\rho]$.
The positivity of the density operator $\rho$ in Eq. (\ref{eq:positivity})
implies that all probabilities are nonnegative: $p_{i}\geq0$. The
completeness equation (\ref{eq:completeness}) together with Eq. (\ref{eq:trace})
implies that the probabilities sum up to one: $\sum_{i}p_{i}=1$.

For a projective measurement, the operators $E_{i}$ are orthogonal
projectors: $E_{i}E_{j}=\delta_{ij}E_{i}$. \emph{\index{Von Neumann!measurement}Von
Neumann measurement} is a special type of a projective measurement,
where the measurement operators $E_{i}$ are orthogonal projectors
with rank one. Such a measurement was considered below Eq. (\ref{eq:spin1/2}),
there the measurement operators are $E_{\uparrow}=\ket{\uparrow}\bra{\uparrow}$
and $E_{\downarrow}=\ket{\downarrow}\bra{\downarrow}$. In general,
the measurement operators do not have to be projectors, they only
need to satisfy the completeness equation (\ref{eq:completeness}). 

For composite systems consisting of two subsystems, Alice and Bob,
it is possible to perform \emph{local measurements} on one of the
subsystems. If a local measurement is done on Alice's subsystem, the
subsystem of Bob remains unchanged. In this case, the measurement
operators have the form $E_{i}=E_{i}^{A}\otimes\openone^{B}$, with
the identity operator $\openone^{B}$ on Bob's Hilbert space. Similarly,
measurement operators corresponding to local measurement on Bob's
subsystem have the form $E_{i}=\openone^{A}\otimes E_{i}^{B}$.

Finally, we also mention the concept of \emph{\index{Quantum!operation}quantum
operations}, which is closely related to quantum measurements. Any
set of measurement operators $\{E_{i}\}$ can also be called a quantum
operation. The corresponding operators $E_{i}$ are then called \index{Kraus operators}\emph{Kraus
operators}. The action of a quantum operation $\{E_{i}\}$ on a density
operator $\rho$ is given by 
\begin{equation}
\Lambda(\rho)=\sum_{i}E_{i}\rho E_{i}^{\dagger}.
\end{equation}
For composite systems, \emph{local quantum operations} can be defined
in the same way as it was done for local measurements. The importance
of quantum operations lies in the fact that they describe the most
general change of a quantum state possible in experiments. Quantum
operations also play an important role in the study of noisy systems:
noise is usually modeled as a quantum operation.

\section{Reduced density operator}

Sometimes one is only interested in one of the subsystems of a composite
quantum system. This situation is captured by the concept of the \emph{reduced
density operator\index{Reduced density operator}}. If the total system
is described by the density operator $\rho^{AB}$, then the system
of $A$ is described by the reduced density operator 
\begin{equation}
\rho^{A}=\mathrm{Tr}_{B}[\rho^{AB}],
\end{equation}
where $\mathrm{Tr}_{B}$ is called \emph{\index{Partial trace}partial
trace} over the subsystem $B$. The partial trace is defined by 
\begin{equation}
\mathrm{Tr}_{B}[\ket{a_{1}}\bra{a_{2}}\otimes\ket{b_{1}}\bra{b_{2}}]=\ket{a_{1}}\bra{a_{2}}\mathrm{Tr}[\ket{b_{1}}\bra{b_{2}}],\label{eq:partial trace}
\end{equation}
where $\ket{a_{1}}$ and $\ket{a_{2}}$ are any two vectors in ${\cal H}_{A}$,
and $\ket{b_{1}}$ and $\ket{b_{2}}$ are any two vectors in ${\cal H}_{B}$.
The trace on the right hand side is the usual trace for the subsystem
$B$: $\mathrm{Tr}[\ket{b_{1}}\bra{b_{2}}]=\braket{b_{2}|b_{1}}$.
In addition to Eq. (\ref{eq:partial trace}), we also require that
the partial trace is linear, i.e., $\mathrm{Tr}_{B}[M^{AB}+N_{}^{AB}]=\mathrm{Tr}_{B}[M^{AB}]+\mathrm{Tr}_{B}[N^{AB}]$
for any two operators $M^{AB}$ and $N^{AB}$. In this way, the partial
trace is defined for all density operators. The physical meaning of
the partial trace lies in the fact that it is the unique operation
for obtaining correct measurement statistics for the subsystem $A$
\citep[p. 105ff.]{Nielsen2000}.

\section{Entropy and mutual information}

\index{Von Neumann!entropy}The \emph{von Neumann entropy} of a quantum
state with density operator $\rho$ is defined as 
\begin{equation}
S(\rho)=-\mathrm{Tr}[\rho\log_{2}\rho],
\end{equation}
where the logarithm of the density operator $\rho$ is defined via
its eigenvalues $\lambda_{i}$ and eigenstates $\ket{i}$ in the following
way: $ $$\log_{2}\rho=\sum_{i}\log_{2}(\lambda_{i})\ket{i}\bra{i}$.
With this definition, the entropy can be written as 
\begin{equation}
S(\rho)=-\sum_{i}\lambda_{i}\log_{2}\lambda_{i},
\end{equation}
where it is defined that $0\log_{2}0=0$. 

The von Neumann entropy is the quantum version of the classical \index{Shannon entropy}\emph{Shannon
entropy}. For a discrete random variable $X$ which can take a value
$x$ with probability $p_{x}$, the Shannon entropy is defined as
\begin{equation}
H(X)=-\sum_{x}p_{x}\log_{2}p_{x}.
\end{equation}
Similar to the Shannon entropy, which measures the uncertainty of
a classical random variable, the von Neumann entropy measures the
uncertainty of a quantum state. Pure states represent full knowledge
about a quantum system: their von Neumann entropy is zero. On the
other hand, for a $d$-dimensional Hilbert space, maximal uncertainty
is represented by the completely mixed density operator $\openone/d$
with the von Neumann entropy $\log_{2}d$.

For two parties, the von Neumann entropy can be used to define the
\emph{mutual information\index{Mutual information}} between the parties.
If the total state is given by the density operator $\rho^{AB}$ with
reduced density operators $\rho^{A}$ and $\rho^{B}$, the mutual
information is defined as 
\begin{equation}
I(\rho^{AB})=S(\rho^{A})+S(\rho^{B})-S(\rho^{AB}).\label{eq:mutual information}
\end{equation}
The mutual information is zero if the state is completely uncorrelated,
i.e., if the density operator has the form $\rho^{AB}=\rho^{A}\otimes\rho^{B}$.
Otherwise, the mutual information is greater than zero: it measures
the amount of correlations between $A$ and $B$.

Closely related to the von Neumann entropy is the \emph{quantum }\index{Relative entropy}\emph{relative
entropy}. For two density operators $\rho$ and $\sigma$ it is defined
as
\begin{equation}
S(\rho||\sigma)=\mathrm{Tr}[\rho\log_{2}\rho]-\mathrm{Tr}[\rho\log_{2}\sigma]\label{eq:relative entropy}
\end{equation}
if the support of $\rho$ is contained in the support of $\sigma$,
and $S(\rho||\sigma)=+\infty$ otherwise. The quantum relative entropy
is nonnegative, and zero if and only if $\rho=\sigma$. The mutual
information defined in Eq. (\ref{eq:mutual information}) can be written
as the relative entropy between the density operator $\rho^{AB}$
and the tensor product of the reduced density operators $\rho^{A}\otimes\rho^{B}$
\citep{Vedral2002}: 
\begin{equation}
I(\rho^{AB})=S(\rho^{AB}||\rho^{A}\otimes\rho^{B}).
\end{equation}

\section{Distance between density operators}

Given two quantum states, how ``close'' are they to each other?
This question, posed in \citep[p. 403]{Nielsen2000}, can be answered
by defining an appropriate distance onto the set of density operators.
One important and frequently used distance is the \emph{trace distance\index{Trace distance}}
\begin{equation}
D_{t}(\rho,\sigma)=\frac{1}{2}\mathrm{Tr}|\rho-\sigma|,
\end{equation}
where $\rho$ and $\sigma$ are any two density operators, $\mathrm{Tr}|M|=\mathrm{Tr}\sqrt{M^{\dagger}M}$
is the trace norm of an operator $M$, and the square root of a Hermitian
operator $M^{\dagger}M$ with nonnegative eigenvalues $\lambda_{i}$
and eigenstates $\ket{i}$ is defined as $\sqrt{M^{\dagger}M}=\sum_{i}\sqrt{\lambda_{i}}\ket{i}\bra{i}$.
The trace distance satisfies all properties of a general distance
$D$: 
\begin{itemize}
\item $D(\rho,\sigma)\geq0$, and $D(\rho,\sigma)=0$ holds if and only
if $\rho=\sigma$,
\item $D$ is symmetric: $D(\rho,\sigma)=D(\sigma,\rho)$,
\item $D$ satisfies the triangle inequality: $D(\rho,\tau)\leq D(\rho,\sigma)+D(\sigma,\tau)$
for any three density operators $\rho$, $\sigma$, and $\tau$.
\end{itemize}
In quantum information theory, the trace distance has an important
interpretation: $\frac{1}{2}+\frac{1}{2}D_{t}(\rho,\sigma)$ is the
optimal probability of success for distinguishing two quantum states
with density operators $\rho$ and $\sigma$ \citep{Fuchs1999}.

Another frequently used quantity is the \emph{fidelity}\index{Fidelity}.
For two density operators $\rho$ and $\sigma$ it is defined as 
\begin{equation}
F(\rho,\sigma)=\left(\mathrm{Tr}\sqrt{\sqrt{\rho}\sigma\sqrt{\rho}}\right)^{2}.
\end{equation}
The fidelity itself is not a distance, since it is one if and only
if $\rho=\sigma$, and smaller than one otherwise. However, the fidelity
can be used to define the \index{Bures distance}\emph{Bures distance}:
$D_{B}(\rho,\sigma)=2(1-\sqrt{F(\rho,\sigma)})$, which satisfies
all properties of a mathematical distance.

Both, the trace distance and the Bures distance have also another
important property, namely they are \emph{nonincreasing under quantum
operations}: 
\begin{equation}
D(\Lambda(\rho),\Lambda(\sigma))\leq D(\rho,\sigma),\label{eq:nonincreasing}
\end{equation}
where $\rho$ and $\sigma$ are any two density operators, and $\Lambda$
is any quantum operation. This property is frequently used in quantum
information theory, especially in studying entanglement and other
quantum correlations.

Note that the inequality (\ref{eq:nonincreasing}) does not follow
from the general properties of a mathematical distance, and thus there
exist distances which violate it. One such distance is the \index{Hilbert-Schmidt distance}\emph{Hilbert-Schmidt
distance} 
\begin{equation}
D_{HS}(\rho,\sigma)=\left\Vert \rho-\sigma\right\Vert ,
\end{equation}
where $\left\Vert M\right\Vert =\sqrt{\mathrm{Tr}[M^{\dagger}M]}$
is the Hilbert-Schmidt norm of an operator $M$. For the Hilbert-Schmidt
distance violation of Eq. (\ref{eq:nonincreasing}) was shown in \citep{Ozawa2000,Piani2012}.

Finally, the relative entropy introduced in Eq. (\ref{eq:relative entropy})
is not a distance in the mathematical sense since it is not symmetric,
and also does not satisfy the triangle inequality. However, the relative
entropy is nonincreasing under quantum operations, i.e., it satisfies
the inequality (\ref{eq:nonincreasing}) \citep{Lindblad1975}.

\chapter{Quantum entanglement}

\section{Definition}

\index{Entanglement} For two parties, Alice ($A$) and Bob ($B$),
the state of the total quantum system can have product form%
\footnote{Sometimes we write $\ket{a}\ket{b}$ or $\ket{ab}$ instead of $\ket{a}\otimes\ket{b}$.%
}: 
\begin{equation}
\ket{\Psi}=\ket{a}\otimes\ket{b},\label{eq:product}
\end{equation}
where the states $\ket{a}$ and $\ket{b}$ are elements of the corresponding
local Hilbert spaces ${\cal H}_{A}$ and ${\cal H}_{B}$. States of
the form given in Eq. (\ref{eq:product}) are not entangled, they
are also called \emph{separable}\index{Separable state}. However,
not all states are separable, since quantum mechanics also allows
superpositions which are not necessarily product: 
\begin{equation}
\ket{\Phi}=\frac{1}{N}(\ket{a_{1}}\otimes\ket{b_{1}}+\ket{a_{2}}\otimes\ket{b_{2}}),
\end{equation}
where $N$ assures normalization such that $\braket{\Phi|\Phi}=1$.
If $\ket{\Phi}$ cannot be written as a product, i.e., $\ket{\Phi}\neq\ket{a}\otimes\ket{b}$,
the state is called \emph{entangled}.
\begin{example*}
The singlet\index{Singlet} state $\ket{\Phi}=\frac{1}{\sqrt{2}}(\ket{01}-\ket{10})$
is entangled, it cannot be written as a product. 
\end{example*}
A mixed state is separable if it can be written as a convex combination
of pure product states \citep{Werner1989}: 
\begin{equation}
\rho_{\mathrm{sep}}=\sum_{i}p_{i}\ket{a_{i}}\bra{a_{i}}\otimes\ket{b_{i}}\bra{b_{i}}.\label{eq:separable}
\end{equation}
The pure states $\ket{a_{i}}$ and $\ket{b_{i}}$ are elements of
the local Hilbert spaces ${\cal H}_{A}$ and ${\cal H}_{B}$, and
$p_{i}\geq0$ are probabilities summing up to one: $\sum_{i}p_{i}=1$.
If the state cannot be written in this form, it is called entangled.

The idea behind this definition of entanglement is the following:
suppose that Alice and Bob are able to produce any quantum state locally.
In addition, they have access to a classical communication channel,
such as a telephone. Then, Alice and Bob can produce any separable
state as given in Eq. (\ref{eq:separable}) by the following procedure:
Alice prepares the state $\ket{a_{i}}$ with the probability $p_{i}$,
and lets Bob know which state she prepared. Depending on this information,
Bob prepares the corresponding state $\ket{b_{i}}$. On the other
hand, it is not possible to create entangled states such as the singlet
state in this way.

\section{Local operations and classical communication}

\index{LOCC} The process for creating separable states presented
above belongs to the class of \emph{local operations and classical
communication} (LOCC), first introduced in \citep{Bennett1996b}.
This class of operations describes the most general procedure Alice
and Bob can apply in quantum theory, if they are limited to classical
communication only. The full mathematical description of these operations
is demanding, and still subject of extensive research \citep{Chitambar2014}.
However, the general idea is simple, and will be explained in the
following.

For two parties, Alice and Bob, a quantum operation $\Lambda_{\mathrm{LOCC}}$
belongs to the class of LOCC, if it can be decomposed into the following
steps: 
\begin{enumerate}
\item \noindent One of the parties, e.g. Alice, performs a local measurement
on her subsystem.
\item \noindent The outcome of the measurement is communicated \emph{classically}
to the other party, here Bob.
\item \noindent Depending on the received information, Bob performs a local
measurement on his subsystem.
\item \noindent The outcome of Bob's measurement is communicated \emph{classically}
to Alice.
\item \noindent Depending on the received information, Alice performs a
local measurement on her subsystem, and the process starts over at
step 2.
\end{enumerate}
The class of LOCC plays an important role in quantum information theory,
especially when studying entanglement. As we have mentioned above,
any separable state can be created with LOCC. On the other hand, LOCC
cannot be used to create entangled states \citep{Horodecki2009}.

\section{Entanglement as a resource}

Until the 1990s, quantum entanglement was mainly regarded as a physical
curiosity: an exotic feature with no practical use. This situation
started to change in 1991, when Ekert presented the first task in
quantum information theory which was based on entanglement \citep{Ekert1991}.
In his work, Ekert showed that if two parties, Alice and Bob, share
a large amount of entangled singlet states, they can communicate in
a completely secure way. This task is referred to as \emph{quantum
cryptography}, or \emph{quantum key distribution}. This strong result
should be compared to the classical cryptography as we use it today.
The security of classical cryptography is mainly based on the conjecture
that a large number is hard to factorize, whereas the quantum cryptography
protocol presented by Ekert is provably secure.

Motivated by Ekert's result, several tasks involving entanglement
have been presented in the following years. In 1992 Bennett and Wiesner
showed that two entangled parties can communicate two classical bits
by sending only one qubit, i.e., one quantum system on a two-dimensional
Hilbert space \citep{Bennett1992}. This task is also known as \emph{quantum
dense coding}, since it suggests that two classical bits can be coded
into one quantum bit.

Another application for entanglement has been proposed in \citep{Bennett1993}.
The authors studied the task of communicating an unknown quantum state
between two parties. An unknown quantum state cannot be communicated
by classical means, which is a direct consequence of the fact that
such a state cannot be cloned \citep{Wootters1982}. However, if the
two parties share an entangled singlet, Bennett \emph{et al.} showed
that any unknown quantum bit can be perfectly communicated. This task
is also known as \emph{quantum teleportation}.

\section{Entanglement measures}

The tasks presented above, namely quantum cryptography, dense coding
and teleportation demonstrate the role of entanglement for a very
special case. In particular, two parties, Alice and Bob, need to share
entangled singlets in order to perform these tasks. However, a pure
quantum state is not necessarily a singlet, and in a realistic scenario
the quantum state is usually mixed. For this reason it is natural
to ask whether a general mixed quantum state can also be used for
some of these tasks. 

The ``usefulness'' of a quantum state for one of the tasks presented
above is usually quantified by the amount of entanglement contained
in the state. One of the most popular quantifiers is the \index{Distillable entanglement}
\emph{distillable entanglement} \citep{Bennett1996a}: it is defined
as the maximal number of singlets that can be obtained per copy of
a given mixed state via local operations and classical communication,
if the number of copies goes to infinity%
\footnote{See also \citep{Horodecki2009} for a formal definition.%
}. The major disadvantage of the distillable entanglement is the fact
that it is hard to evaluate. Thus, exact expressions are only known
in a few special cases. For this reason, other quantifiers, known
as \index{Entanglement!measure} \emph{entanglement measures}, have
been proposed in the literature. Any entanglement measure $E$ fulfills
the following two properties \citep{Horodecki2009}:
\begin{enumerate}
\item $E$ does not increase under local operations and classical communication,
\item $E$ vanishes on separable states.\label{entanglement}
\end{enumerate}
For a \emph{pure state} $\ket{\psi}^{AB}$ distributed between two
parties, Alice and Bob, entanglement is usually quantified by the
von Neumann entropy of the reduced density operator $\rho^{A}=\mathrm{Tr}_{B}[\ket{\psi}\bra{\psi}^{AB}]$:
\begin{equation}
E(\ket{\psi}^{AB})=S(\rho^{A})=-\sum_{i}\lambda_{i}\log_{2}\lambda_{i},\label{eq:entanglement entropy}
\end{equation}
where $\lambda_{i}$ are the eigenvalues of $\rho^{A}$. The importance
of this quantity in quantum information theory comes from the fact
that it is equal to the distillable entanglement for all pure states
\citep{Bennett1996}.

For a \emph{mixed state} $\rho^{AB}$, two main classes of entanglement
measures are considered in the literature. These are
\begin{itemize}
\item convex roof measures and
\item distance-based measures.
\end{itemize}
Any measure of entanglement $E$ which is defined on all pure states
can be extended to mixed states via the following \emph{convex roof}
construction \citep{Uhlmann1998}: 
\begin{equation}
E(\rho)=\inf_{\left\{ p_{i},\ket{\psi_{i}}\right\} }\sum_{i}p_{i}E(\ket{\psi_{i}}),\label{eq:convex roof}
\end{equation}
where the infimum is taken over all decompositions $\left\{ p_{i},\ket{\psi_{i}}\right\} $
of the given density operator $\rho$ with nonnegative probabilities
$p_{i}$, i.e., $\rho=\sum_{i}p_{i}\ket{\psi_{i}}\bra{\psi_{i}}$. 

For bipartite systems, the \emph{\index{Entanglement!of formation}}
\emph{entanglement of formation} defined in \citep{Bennett1996b}
is one of the most popular and frequently used convex roof measures.
For pure states it is defined as the von Neumann entropy of the reduced
density operator in Eq. (\ref{eq:entanglement entropy}). The extension
to mixed states is done via the convex roof construction in Eq. (\ref{eq:convex roof}).
Although the infimum in Eq. (\ref{eq:convex roof}) is hard to evaluate
in general, Wootters presented a closed expression for the entanglement
of formation for all mixed states of two qubits \citep{Wootters1998}.
For any such state, the entanglement of formation $E_{f}$ is given
by 
\begin{equation}
E_{f}(\rho)=h\left(\frac{1}{2}+\frac{1}{2}\sqrt{1-C^{2}(\rho)}\right)
\end{equation}
with the binary entropy $h(x)=-x\log_{2}x-(1-x)\log_{2}(1-x)$, and
the \index{Concurrence} concurrence $C(\rho)=\max\{0,\lambda_{1}-\lambda_{2}-\lambda_{3}-\lambda_{4}\}$,
where $\lambda_{i}$ are the square roots of the eigenvalues of $\rho\tilde{\rho}$
in decreasing order, and $\tilde{\rho}$ is defined as $\tilde{\rho}=(\sigma_{y}\otimes\sigma_{y})\rho^{*}(\sigma_{y}\otimes\sigma_{y})$
with the Pauli matrix $\sigma_{y}=\left(\begin{array}{cc}
0 & -i\\
i & 0
\end{array}\right)$. The entanglement of formation satisfies the criteria for a proper
entanglement measure given on page \pageref{entanglement}: it does
not increase under local operations and classical communication and
vanishes on separable states. While the second property is easy to
verify, the first property was proven in \citep{Bennett1996b}.

The second main class of entanglement measures are measures based
on distance proposed in \citep{Vedral1997}. All those measures can
be written as 
\begin{equation}
E(\rho)=\inf_{\sigma\in{\cal S}}D(\rho,\sigma),
\end{equation}
where $D$ is a distance, and the infimum is taken over the set of
separable states ${\cal S}$. If the distance $D$ does not increase
under quantum operations, i.e., 
\begin{equation}
D(\Lambda(\rho),\Lambda(\sigma))\leq D(\rho,\sigma)\label{eq:contractive}
\end{equation}
for any quantum operation $\Lambda$ and any two states $\rho$ and
$\sigma$, then the corresponding measure of entanglement does not
increase under local operations and classical communication \citep{Vedral1997}.
This property is satisfied by the relative entropy $S(\rho||\sigma)=\mathrm{Tr}[\rho\log_{2}\rho]-\mathrm{Tr}[\rho\log_{2}\sigma]$,
although the relative entropy is not a distance in the mathematical
sense. The corresponding measure of entanglement is called \index{Relative entropy!of entanglement}
\emph{relative entropy of entanglement}: 
\begin{equation}
E_{R}(\rho)=\min_{\sigma\in{\cal S}}S(\rho||\sigma).
\end{equation}
The relative entropy of entanglement is one of the most popular and
widely studied measures of entanglement. One reason is the fact that
the relative entropy itself plays an important role in quantum information
theory \citep{Vedral2002}. Moreover, the relative entropy of entanglement
is a powerful upper bound for the distillable entanglement \citep{Horodecki2000}.

We have already mentioned above that all distance-based entanglement
measures do not increase under local operations and classical communication,
if the distance satisfies Eq. (\ref{eq:contractive}). This is one
of the properties any reasonable measure of entanglement should satisfy.
Moreover, any entanglement measure should also vanish on separable
states. This is also easily seen to be true for any distance $D(\rho,\sigma)$
which is zero if and only if $\rho=\sigma$, and larger than zero
otherwise.

Finally, we mention the relation between three of the measures presented
in this section, namely between the distillable entanglement $E_{d}$,
the relative entropy of entanglement $E_{R}$, and the entanglement
of formation $E_{f}$. As was shown in \citep{Horodecki2000}, these
measures satisfy the inequality 
\begin{equation}
E_{d}\leq E_{R}\leq E_{f}
\end{equation}
for all mixed states, i.e., the relative entropy of entanglement is
always between $E_{d}$ and $E_{f}$.

\chapter{Quantum correlations beyond entanglement}

\section{Definition }

\index{Quantum!correlations}

A mixed state shared by two parties, Alice and Bob, is called \emph{classically
correlated\index{Classical correlations}} if it can be written as
\citep{Oppenheim2002} 
\begin{equation}
\rho_{\mathrm{cc}}=\sum_{i,j}p_{ij}\ket{i}\bra{i}^{A}\otimes\ket{j}\bra{j}^{B},
\end{equation}
where $\{\ket{i}^{A}\}$ are orthogonal states on Alice's Hilbert
space ${\cal H}_{A}$ and $\{\ket{j}^{B}\}$ are orthogonal states
on Bob's Hilbert space ${\cal H}_{B}$. The probabilities $p_{ij}$
are nonnegative and sum up to one: $\sum_{i,j}p_{ij}=1$. Otherwise
the state is called \emph{quantum correlated}. Note that every classically
correlated state is also separable. On the other hand, a separable
state $\rho_{\mathrm{sep}}=\sum_{i}p_{i}\ket{a_{i}}\bra{a_{i}}\otimes\ket{b_{i}}\bra{b_{i}}$
is not necessarily classically correlated, since the states $\{\ket{a_{i}}\}$
and $\{\ket{b_{i}}\}$ do not have to be orthogonal. Moreover, a pure
state is quantum correlated if and only if the state is entangled,
i.e., both concepts are equivalent for pure states. For this reason,
we will discuss mixed states in the following.

The intuition behind this definition of classically correlated states
comes from the fact that these states are not disturbed by certain
local von Neumann measurements on Alice's and Bob's subspaces. The
measurement operators corresponding to these non-disturbing von Neumann
measurements are given by $E_{i}^{A}=\ket{i}\bra{i}^{A}$ and $E_{j}^{B}=\ket{j}\bra{j}^{B}$.
In a similar way we can also define a class of quantum states which
is not disturbed under certain von Neumann measurements on the subspace
of one party (e.g. Alice) only. In this case the state has the form
\begin{equation}
\rho_{\mathrm{cq}}=\sum_{i}p_{i}\ket{i}\bra{i}^{A}\otimes\rho_{i}^{B},
\end{equation}
where $\ket{i}^{A}$ are orthogonal states on Alice's Hilbert space
${\cal H}_{A}$, $\rho_{i}^{B}$ are states on Bob's Hilbert space
${\cal H}_{B}$, and the nonnegative probabilities $p_{i}$ sum up
to one. These states are called \index{Classical-quantum states}\emph{classical-quantum}
states \citep{Horodecki2005,Piani2008}. The corresponding von Neumann
measurement on Alice's subsystem which does not disturb the total
state is given by the measurement operators $E_{i}^{A}=\ket{i}\bra{i}^{A}$.
Similarly, a \emph{quantum-classical} state has the form $\rho_{\mathrm{qc}}=\sum_{i}p_{i}\rho_{i}^{A}\otimes\ket{i}\bra{i}^{B}$.
Such a state is not disturbed by a local von Neumann measurement on
Bob's subspace with measurement operators $E_{i}^{B}=\ket{i}\bra{i}^{B}$.

\section{Measures of quantum correlations}

A measure of entanglement can be defined via the usefulness of a quantum
state to perform certain tasks. The figure of merit is the distillable
entanglement, which quantifies how many singlets can be extracted
per copy of a given quantum state via local operations and classical
communication, if many copies of the same state are available. Since
singlets can be used for many tasks in quantum information theory,
e.g., quantum cryptography, dense coding and teleportation, the distillable
entanglement is directly related to the performance of these tasks.

For general quantum correlations the situation is less clear, since
the definition of ``distillable quantum correlations'' is meaningless,
at least if the concept of local operations and classical communication
is considered. The reason for this is the fact that local operations
and classical communication can be used to create an arbitrary amount
of quantum correlations \citep{Dakic2010,Streltsov2011a}. This means
that a measure of ``distillable quantum correlations'' would be
infinite for all quantum states. However, several other approaches
to quantify quantum correlations have been proposed in the literature.
The most important measures of quantum correlations will be presented
in the following.

\subsection{Quantum discord}

\emph{\index{Discord}}Quantum discord is historically the first measure
of quantum correlations beyond entanglement \citep{Zurek2000,Ollivier2001,Henderson2001}.
The definition of quantum discord is based on the fact that in classical
information theory the mutual information between two random variables
$X$ and $Y$ can be expressed in two different ways, namely 
\begin{eqnarray}
I(X:Y) & = & H(X)+H(Y)-H(X,Y),\nonumber \\
J(X:Y) & = & H(X)-H(X|Y).
\end{eqnarray}
Here, $H(X)=-\sum_{x}p_{x}\log_{2}p_{x}$ is the classical Shannon
entropy of the random variable $X$, where $p_{x}$ is the probability
that the random variable $X$ takes the value $x$. $H(X,Y)$ is the
joint entropy of both variables $X$ and $Y$. The conditional entropy
$H(X|Y)$ is defined as 
\begin{equation}
H(X|Y)=\sum_{y}p_{y}H(X|y),
\end{equation}
where $p_{y}$ is the probability that the random variable $Y$ takes
the value $y$, and $H(X|y)$ is the entropy of the variable $X$
conditioned on the variable $Y$ taking the value $y$: $H(X|y)=-\sum_{x}p_{x|y}\log_{2}p_{x|y}$,
and $p_{x|y}$ is the probability of $x$ given $y$.

The equality of $I$ and $J$ for classical random variables follows
from Bayes' rule $p_{x|y}=p_{xy}/p_{y}$, which can be used to show
that $H(X|Y)=H(X,Y)-H(Y)$. However, as was noticed in \citep{Ollivier2001},
$I$ and $J$ are no longer equal if quantum theory is applied. In
particular, for a quantum state $\rho^{AB}$ the mutual information
between $A$ and $B$ is given by 
\begin{equation}
I(\rho^{AB})=S(\rho^{A})+S(\rho^{B})-S(\rho^{AB})
\end{equation}
with the von Neumann entropy $S$, and the reduced density operators
$\rho^{A}=\mathrm{Tr}_{B}[\rho^{AB}]$ and $\rho^{B}=\mathrm{Tr}_{A}[\rho^{AB}]$.
This expression is the generalization of the classical mutual information
$I(X:Y)$ to the quantum theory. 

On the other hand, the generalization of $J(X:Y)$ is not completely
straightforward. Ollivier and Zurek have proposed the following way
to generalize $J$ to the quantum theory \citep{Ollivier2001}: for
a bipartite quantum state $\rho^{AB}$ they defined the conditional
entropy of $A$ conditioned on a measurement on $B$: 
\begin{equation}
S(A|\{\Pi_{i}^{B}\})=\sum_{i}p_{i}S(\rho_{i}^{A}),
\end{equation}
where $\{\Pi_{i}^{B}\}$ are measurement operators corresponding to
a von Neumann measurement on the subsystem $B$, i.e., orthogonal
projectors with rank one. The probability $p_{i}$ for obtaining the
outcome $i$ is given by $p_{i}=\mathrm{Tr}[\Pi_{i}^{B}\rho^{AB}]$,
and the corresponding post-measurement state of the subsystem $A$
is given by $\rho_{i}^{A}=\mathrm{Tr}_{B}[\Pi_{i}^{B}\rho^{AB}]/p_{i}$.
The quantity $J$ can now be extended to quantum states as follows
\citep{Ollivier2001}: 
\begin{equation}
J(\rho^{AB})_{\{\Pi_{i}^{B}\}}=S(\rho^{A})-S(A|\{\Pi_{i}^{B}\}),\label{eq:J}
\end{equation}
where the index $\{\Pi_{i}^{B}\}$ clarifies that the value depends
on the choice of the measurement operators $\Pi_{i}^{B}$. The quantity
$J$ represents the amount of information gained about the subsystem
$A$ by measuring the subsystem $B$ \citep{Ollivier2001}.

\emph{Quantum discord} is the difference of these two inequivalent
expressions for the mutual information, minimized over all von Neumann
measurements: 
\begin{equation}
\delta^{B|A}(\rho^{AB})=\min_{\{\Pi_{i}^{B}\}}\left[I(\rho^{AB})-J(\rho^{AB})_{\{\Pi_{i}^{B}\}}\right],
\end{equation}
where the minimum over all von Neumann measurements is taken in order
to have a measurement-independent expression \citep{Ollivier2001}.
As was also shown in \citep{Ollivier2001}, quantum discord is nonnegative,
and is equal to zero on quantum-classical states only. These are states
of the form $\rho_{\mathrm{qc}}=\sum_{i}p_{i}\rho_{i}^{A}\otimes\ket{i}\bra{i}^{B}$.

A closely related quantity was proposed by Henderson and Vedral in
\citep{Henderson2001}. The authors aimed to quantify \emph{classical
correlations\index{Classical correlations}} in quantum states by
defining a measure of classical correlations $C_{B}$ which is equal
to $J$ given in Eq. (\ref{eq:J}), maximized over all positive operator-valued
measures (POVMs) on the subsystem $B$: 
\begin{equation}
C_{B}(\rho^{AB})=\sup_{\{M_{i}^{B}\}}J(\rho^{AB})_{\{M_{i}^{B}\}}.
\end{equation}
Here, $M_{i}^{B}$ are POVM elements on the subsystem $B$, and $J(\rho^{AB})_{\{M_{i}^{B}\}}$
is the generalization of Eq. (\ref{eq:J}) to POVMs: 
\begin{equation}
J(\rho^{AB})_{\{M_{i}^{B}\}}=S(\rho^{A})-S(A|\{M_{i}^{B}\})
\end{equation}
with $S(A|\{M_{i}^{B}\})=\sum_{i}p_{i}S(\rho_{i}^{A})$. The measurement
probabilities are now given by $p_{i}=\mathrm{Tr}[M_{i}^{B}\rho^{AB}]$,
and the corresponding post-measurement state of the subsystem $A$
is given by $\rho_{i}^{A}=\mathrm{Tr}_{B}[M_{i}^{B}\rho^{AB}]/p_{i}$.

In today's literature, quantum discord is frequently defined as the
difference between the mutual information $I$, and the amount of
classical correlations $C_{B}$ \citep{Datta2008}: 
\begin{equation}
D^{B|A}(\rho^{AB})=I(\rho^{AB})-C_{B}(\rho^{AB}).\label{eq:discord}
\end{equation}
This measure is in general different from the original quantum discord
$\delta^{B|A}$ proposed by Ollivier and Zurek. However, this quantity
is also nonnegative, and vanishes on quantum-classical states only
\citep{Datta2010}. Quantum discord as defined in Eq. (\ref{eq:discord})
is related to the entanglement of formation $E_{f}$ via the Koashi-Winter
relation\index{Koashi-Winter relation} \citep{Koashi2004,Fanchini2011}:
\begin{equation}
D^{B|A}(\rho^{AB})=E_{f}(\rho^{AC})-S(\rho^{AB})+S(\rho^{B}),
\end{equation}
where the total state $\rho^{ABC}$ is pure, i.e., $\rho^{ABC}=\ket{\psi}\bra{\psi}^{ABC}$.

\subsection{General measures of quantum correlations}

Postulates for general measures of quantum correlations have been
proposed in \citep{Brodutch2012}. There the authors identify three
necessary conditions every measure of quantum correlations $Q$ should
satisfy. These conditions are: 
\begin{enumerate}
\item $Q$ is nonnegative,
\item $Q$ is invariant under local unitary operations,
\item $Q$ is zero on classically correlated states.
\end{enumerate}
Note that both versions of quantum discord, $\delta$ and $D$, satisfy
all these criteria. In the following we will present main measures
of general quantum correlations apart from quantum discord.

\emph{\index{Information deficit}Information deficit} is a measure
of quantum correlations which was originally based on the task of
extracting work from a heat bath using a quantum state \citep{Oppenheim2002,Horodecki2005}.
In particular, the amount of extractable work from a heat bath of
temperature $T$ using a mixed state $\rho$ of $n$ qubits is given
by 
\begin{equation}
W=kT\{n-S(\rho)\},
\end{equation}
where $k$ is the Boltzmann constant and $S$ is the von Neumann entropy.
However, if the state is shared by two parties, Alice and Bob, each
of them having access to the local subsystem only, the amount of extractable
work $W'$ will in general be different from $W$. If Alice is allowed
to perform a single von Neumann measurement on her local system and
send the resulting state to Bob, the maximal amount of work which
Bob can extract from the resulting state in this way is given by 
\begin{equation}
W'=W-kT\cdot\Delta^{A|B}(\rho^{AB}),
\end{equation}
where $\Delta^{A|B}$ is known as the \emph{one-way information deficit}
\citep{Horodecki2005}: 
\begin{equation}
\Delta^{A|B}(\rho^{AB})=\min_{\{\Pi_{i}^{A}\}}S(\rho^{AB}||\sum_{i}\Pi_{i}^{A}\rho^{AB}\Pi_{i}^{A}).
\end{equation}
$S(\rho||\sigma)$ is the relative entropy between the states $\rho$
and $\sigma$, and the minimum is taken over local von Neumann measurements
$\{\Pi_{i}^{A}\}$ on the subsystem $A$. The one-way information
deficit is zero on classical-quantum states only, and can also be
written as the minimal relative entropy between the given state $\rho^{AB}$
and the set of classical-quantum states $CQ$ \citep{Modi2010}: 
\begin{equation}
\Delta^{A|B}(\rho^{AB})=\min_{\sigma^{AB}\in CQ}S(\rho^{AB}||\sigma^{AB}).
\end{equation}
For this reason, this quantity is also called \emph{relative entropy
of discord\index{Relative entropy!of discord}}. In a similar way,
it is possible to define the \emph{relative entropy of quantumness\index{Relative entropy!of quantumness}}
as the minimal relative entropy between $\rho^{AB}$ and the set of
classically correlated states $CC$ \citep{Piani2011}: 
\begin{equation}
Q_{R}(\rho^{AB})=\min_{\sigma^{AB}\in CC}S(\rho^{AB}||\sigma^{AB}).
\end{equation}

Inspired by the expression for the relative entropy of discord as
the minimal relative entropy between a given state and the set of
classical-quantum states $CQ$, Daki\'c \emph{et al.} defined the
\emph{\index{Geometric measure of discord}geometric measure of discord}
as the minimal squared Hilbert-Schmidt distance between a given state
$\rho^{AB}$ and $CQ$ \citep{Dakic2010}: 
\begin{equation}
D_{G}^{A|B}(\rho^{AB})=\min_{\sigma^{AB}\in CQ}\left\Vert \rho^{AB}-\sigma^{AB}\right\Vert ^{2}
\end{equation}
with the Hilbert-Schmidt norm $\left\Vert M\right\Vert =\sqrt{\mathrm{Tr}[M^{\dagger}M]}$.
The main advantage of the geometric measure of discord was already
presented in the original work by Daki\'c \emph{et al}.: this measure
has an analytical expression for all two-qubit states \citep{Dakic2010}.
If $\rho^{AB}$ is a two-qubit state, then the geometric measure of
discord can be written as \citep{Dakic2010}
\begin{equation}
D_{G}^{A|B}(\rho^{AB})=\frac{1}{4}(\boldsymbol{a}^{2}+\mathrm{Tr}[E^{T}E]-k_{\max}),
\end{equation}
where $\boldsymbol{a}$ is a $3$-dimensional vector with entries
$a_{i}=\mathrm{Tr}[(\sigma_{i}\otimes\openone)\rho^{AB}]$, and $E$
is the $3\times3$ correlation tensor with components $E_{ij}=\mathrm{Tr}[(\sigma_{i}\otimes\sigma_{j})\rho^{AB}]$.
The Pauli operators $\sigma_{i}$ are given as $\sigma_{1}=\left(\begin{array}{cc}
0 & 1\\
1 & 0
\end{array}\right)$, $\sigma_{2}=\left(\begin{array}{cc}
0 & -i\\
i & 0
\end{array}\right)$, and $\sigma_{3}=\left(\begin{array}{cc}
1 & 0\\
0 & -1
\end{array}\right)$. Finally, $k_{\max}$ is the largest eigenvalue of the real matrix
$\boldsymbol{a}\boldsymbol{a}^{T}+EE^{T}$.

\chapter{Quantum discord in quantum information theory}

\section{Remote state preparation}

\subsection{Deterministic remote state preparation}

\begin{figure}
\noindent \begin{centering}
\includegraphics[width=1\columnwidth]{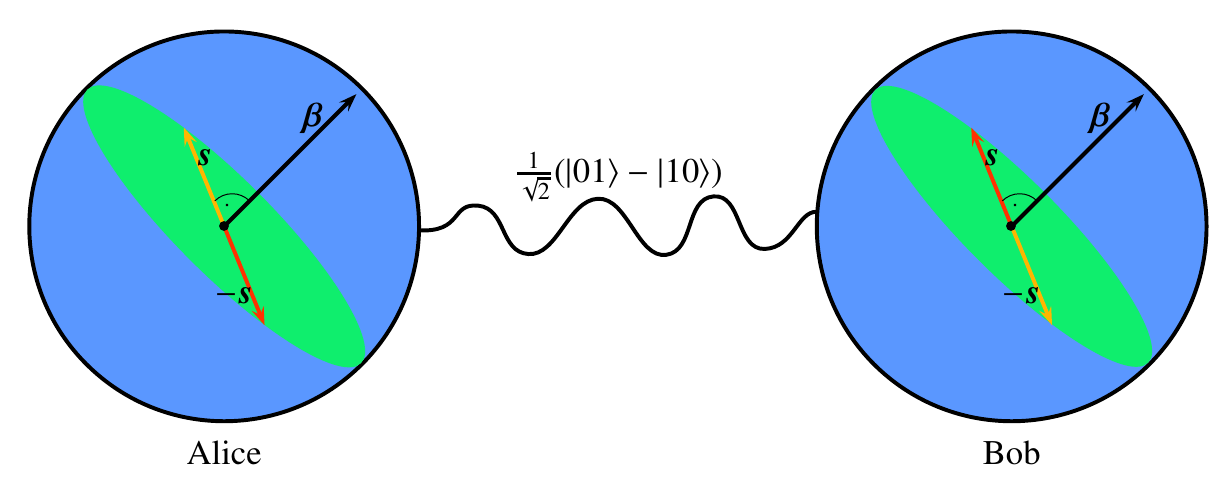}
\par\end{centering}

\caption{\label{fig:RSP}Remote state preparation. Alice can remotely prepare
any state $\ket{\boldsymbol{s}}\bra{\boldsymbol{s}}$ with Bloch vector
$\boldsymbol{s}$ on a fixed equatorial plane of the Bloch sphere
by performing a von Neumann measurement in the basis $\left\{ \ket{-\boldsymbol{s}}\bra{-\boldsymbol{s}},\ket{\boldsymbol{s}}\bra{\boldsymbol{s}}\right\} $
on her part of the singlet and by sending the outcome of the measurement
to Bob. Depending on the outcome, Bob's system is found in one of
the states $\ket{\boldsymbol{s}}\bra{\boldsymbol{s}}$ or $\ket{-\boldsymbol{s}}\bra{-\boldsymbol{s}}$.
Bob can correct the latter by applying a $\pi$ rotation around the
direction $\boldsymbol{\beta}$ orthogonal to the corresponding equatorial
plane.}
\end{figure}
The role of quantum discord in the task of \emph{remote state preparation
\index{Remote state preparation}} was considered by Daki\'c \emph{et
al}.  \citep{Dakic2012}. In this task Alice aims to remotely prepare
Bob's system in the quantum state 
\begin{equation}
\ket{\psi}=\frac{1}{\sqrt{2}}(\ket{0}+e^{i\phi}\ket{1}).\label{eq:RSP}
\end{equation}
To this end Alice and Bob have access to an additional shared quantum
state and a classical communication channel. In contrast to the standard
quantum teleportation \citep{Bennett1993}, which can be applied to
remotely prepare an arbitrary quantum state by making use of a shared
singlet and two bits of classical communication, remote preparation
of the state given in Eq. (\ref{eq:RSP}) requires a shared singlet
supported by only one classical bit \citep{Bennett2001}. To achieve
this task, Alice applies a von Neumann measurement in the basis $\left\{ \ket{\psi_{\bot}}\bra{\psi_{\bot}},\ket{\psi}\bra{\psi}\right\} $
on her part of the singlet, where the state $\ket{\psi_{\bot}}=(\ket{0}-e^{i\phi}\ket{1})/\sqrt{2}$
is orthogonal to $\ket{\psi}$. Depending on her outcome, Bob's system
is found in one of the states $\ket{\psi}\bra{\psi}$ or $\ket{\psi_{\bot}}\bra{\psi_{\bot}}$.
By sending the outcome of her measurement to Bob -- which implies
sending one classical bit -- he either finds his system in the desired
state $\ket{\psi}\bra{\psi}$, or can correct his state $\ket{\psi_{\bot}}\bra{\psi_{\bot}}$
by applying the Pauli operator $\sigma_{z}$. 

Note that the Bloch vector of the state $\ket{\psi}$ in Eq. (\ref{eq:RSP})
lies in the equatorial plane of the Bloch sphere orthogonal to the
z axis. Moreover, the $\sigma_{z}$ operation which is applied by
Bob to the state $\ket{\psi_{\bot}}\bra{\psi_{\bot}}$ can be regarded
as a $\pi$ rotation around the z axis of the corresponding Bloch
vector. In a similar way Alice can remotely prepare any pure state
in a fixed equatorial plane of the Bloch sphere. If $\boldsymbol{s}$
is the Bloch vector of the state $\ket{\boldsymbol{s}}\bra{\boldsymbol{s}}$
Alice wishes to prepare and $\boldsymbol{\beta}$ is a normalized
vector orthogonal to the corresponding equatorial plane, Alice can
achieve this task by performing a von Neumann measurement in the basis
$\left\{ \ket{-\boldsymbol{s}}\bra{-\boldsymbol{s}},\ket{\boldsymbol{s}}\bra{\boldsymbol{s}}\right\} $
on her part of the singlet and send the outcome to Bob. Depending
on the measurement outcome, Bob either finds his system in the state
$\ket{\boldsymbol{s}}\bra{\boldsymbol{s}}$ or $\ket{-\boldsymbol{s}}\bra{-\boldsymbol{s}}$
and can correct the latter by applying a $\pi$ rotation around the
direction $\boldsymbol{\beta}$, see also Fig. \ref{fig:RSP} for
illustration.

\subsection{Remote state preparation in the presence of noise}

So far we considered deterministic remote state preparation where
Alice could remotely prepare the desired state with certainty. However,
this is not possible in general if the state shared by Alice and Bob
is mixed, and the above procedure will leave Bob's system in a mixed
state with Bloch vector $\boldsymbol{r}$. The aim of Alice in this
case is to adjust her measurement such that Bob's final Bloch vector
$\boldsymbol{r}$ becomes as close as possible to the desired vector
$\boldsymbol{s}$. As a quantifier of the performance of this procedure
Daki\'c \emph{et al}. introduced the payoff-function \citep{Dakic2012}
\begin{equation}
{\cal P}=(\boldsymbol{r}\cdot\boldsymbol{s})^{2}.
\end{equation}
For a pure state $\ket{\boldsymbol{s}}$ with Bloch vector $\boldsymbol{s}$
and a mixed state $\rho$ with Bloch vector $\boldsymbol{r}$ the
payoff function is directly related to the fidelity $\braket{\boldsymbol{s}|\rho|\boldsymbol{s}}$
between the two states. This can be seen by writing the fidelity explicitly
as $\braket{\boldsymbol{s}|\rho|\boldsymbol{s}}=(1+\boldsymbol{r}\cdot\boldsymbol{s})/2$,
and thus we get the desired relation ${\cal P}=(2\braket{\boldsymbol{s}|\rho|\boldsymbol{s}}-1)^{2}$.

The aim of Alice is to maximize the payoff function for a given Bloch
vector $\boldsymbol{s}$. As was shown in \citep{Dakic2012}, this
maximization can be performed for any mixed two-qubit state shared
by Alice and Bob. Note that any such state admits the following representation:
\begin{equation}
\rho=\frac{1}{4}\left(\openone\otimes\openone+\sum_{i=1}^{3}a_{i}\cdot\sigma_{i}\otimes\openone+\sum_{j=1}^{3}b_{j}\cdot\openone\otimes\sigma_{j}+\sum_{k,l=1}^{3}E_{kl}\cdot\sigma_{k}\otimes\sigma_{l}\right).\label{eq:two qubits}
\end{equation}
The vectors $\boldsymbol{a}=(a_{1},a_{2},a_{3})$ and $\boldsymbol{b}=(b_{1},b_{2},b_{3})$
are the Bloch vectors of Alice's and Bob's local state respectively,
and $E_{kl}=\mathrm{Tr}[\sigma_{k}\otimes\sigma_{l}\rho]$ are the
elements of the correlation tensor $E$. If Alice applies a von Neumann
measurement in the basis $\left\{ \ket{\boldsymbol{\alpha}}\bra{\boldsymbol{\alpha}},\ket{-\boldsymbol{\alpha}}\bra{-\boldsymbol{\alpha}}\right\} $
on her part of the mixed state, she obtains one of two outcomes with
corresponding probabilities $p_{\boldsymbol{\alpha}}$ and $p_{-\boldsymbol{\alpha}}=1-p_{\boldsymbol{\alpha}}$.
By using the relation $\ket{\boldsymbol{\alpha}}\bra{\boldsymbol{\alpha}}=\frac{1}{2}(\openone+\boldsymbol{\alpha}\cdot\boldsymbol{\sigma})$,
where $\boldsymbol{\sigma}=(\sigma_{1},\sigma_{2},\sigma_{3})$ is
a vector containing the Pauli matrices, we can express the probability
$p_{\boldsymbol{\alpha}}$ as follows: 
\begin{eqnarray}
p_{\boldsymbol{\alpha}} & = & \mathrm{Tr}\left[\ket{\boldsymbol{\alpha}}\bra{\boldsymbol{\alpha}}\otimes\openone\rho\right]=\frac{1}{2}(1+\boldsymbol{\alpha}\cdot\boldsymbol{a}).\label{eq:p}
\end{eqnarray}
Conditioned on the outcome of this measurement, the state of Bob's
system is found to be 
\begin{eqnarray}
\rho_{\boldsymbol{\alpha}}^{B} & = & \frac{\mathrm{Tr}_{A}\left[\ket{\boldsymbol{\alpha}}\bra{\boldsymbol{\alpha}}\otimes\openone\rho\right]}{p_{\boldsymbol{\alpha}}}.\label{eq:rhoB}
\end{eqnarray}
Recall that this state can also be written in the form $\rho_{\boldsymbol{\alpha}}^{B}=(\openone+\boldsymbol{b}_{\boldsymbol{\alpha}}\cdot\boldsymbol{\sigma})/2$
with the Bloch vector $\boldsymbol{b}_{\boldsymbol{\alpha}}$. By
inserting Eqs. (\ref{eq:two qubits}) and (\ref{eq:p}) into Eq. (\ref{eq:rhoB})
the Bloch vector $\boldsymbol{b}_{\boldsymbol{\alpha}}$ can be written
explicitly as 
\begin{eqnarray}
\boldsymbol{b}_{\boldsymbol{\alpha}} & = & \frac{\boldsymbol{b}+E^{T}\boldsymbol{\alpha}}{1+\boldsymbol{\alpha}\cdot\boldsymbol{a}},\label{eq:b}
\end{eqnarray}
where $E^{T}$ is the transposed correlation tensor $E$.

In the next steps Alice and Bob follow the same procedure as for the
deterministic remote state preparation discussed above (see Fig. \ref{fig:RSP}).
In particular, Alice sends the outcome of her measurement to Bob,
who applies a $\pi$ rotation $R_{\pi}$ around the direction $\boldsymbol{\beta}$
conditioned on the outcome of the measurement. After these steps the
Bloch vector $\boldsymbol{r}_{\boldsymbol{\alpha}}$ of Bob's final
state takes the form 
\begin{equation}
\boldsymbol{r}_{\boldsymbol{\alpha}}=p_{\boldsymbol{\alpha}}\boldsymbol{b}_{\boldsymbol{\alpha}}+p_{-\boldsymbol{\alpha}}R_{\pi}\boldsymbol{b}_{-\boldsymbol{\alpha}}.
\end{equation}
Note that this procedure is optimal if the state shared by Alice and
Bob is a singlet, and if Alice chooses her measurement basis $\left\{ \ket{\boldsymbol{\alpha}}\bra{\boldsymbol{\alpha}},\ket{-\boldsymbol{\alpha}}\bra{-\boldsymbol{\alpha}}\right\} $
such that $\boldsymbol{\alpha}=-\boldsymbol{s}$, where $\boldsymbol{s}$
is the Bloch vector of the state $\ket{\boldsymbol{s}}$ Alice wishes
to prepare. In this case the Bloch vector of Bob's final state $\boldsymbol{r}_{-\boldsymbol{s}}$
is equal to $\boldsymbol{s}$.

For evaluating the payoff function ${\cal P}=(\boldsymbol{r}\cdot\boldsymbol{s})^{2}$
it is crucial to note that any vector $\boldsymbol{x}$ satisfies
the following equality: $(R_{\pi}\boldsymbol{x})\cdot\boldsymbol{s}=-\boldsymbol{x}\cdot\boldsymbol{s}$.
This equality can be proven by using the invariance of the scalar
product under rotations, i.e., $ $ $(R_{\pi}\boldsymbol{x})\cdot(R_{\pi}\boldsymbol{y})=\boldsymbol{x}\cdot\boldsymbol{y}$
for any two vectors $\boldsymbol{x}$ and $\boldsymbol{y}$. With
this in mind, we see that $(R_{\pi}\boldsymbol{x})\cdot\boldsymbol{s}=(R_{\pi}^{2}\boldsymbol{x})\cdot(R_{\pi}\boldsymbol{s})=-\boldsymbol{x}\cdot\boldsymbol{s}$,
where in the last step we used the fact that double application of
the rotation $R_{\pi}$ does not change the vector, i.e., $R_{\pi}^{2}\boldsymbol{x}=\boldsymbol{x}$,
and that the rotation $R_{\pi}$ applied to the vector $\boldsymbol{s}$
takes it to $-\boldsymbol{s}$, see Fig. \ref{fig:RSP}. Using these
results the product $\boldsymbol{r}\cdot\boldsymbol{s}$ takes the
following form: $\boldsymbol{r}\cdot\boldsymbol{s}=p_{\boldsymbol{\alpha}}\boldsymbol{b}_{\boldsymbol{\alpha}}\cdot\boldsymbol{s}-p_{-\boldsymbol{\alpha}}\boldsymbol{b}_{-\boldsymbol{\alpha}}\cdot\boldsymbol{s}$.
By using Eqs. (\ref{eq:p}) and (\ref{eq:b}) this product can also
be written as $\boldsymbol{r}\cdot\boldsymbol{s}=\boldsymbol{\alpha}E\boldsymbol{s}$,
and the payoff function reduces to 
\begin{equation}
\mathcal{P}=(\boldsymbol{\alpha}E\boldsymbol{s})^{2}.
\end{equation}
This expression is valid for any von Neumann measurement of Alice
in the basis $\left\{ \ket{\boldsymbol{\alpha}}\bra{\boldsymbol{\alpha}},\ket{-\boldsymbol{\alpha}}\bra{-\boldsymbol{\alpha}}\right\} $.
Maximal payoff is achieved if the vector $\boldsymbol{\alpha}$ is
parallel to the vector $E\boldsymbol{s}$, i.e., $\boldsymbol{\alpha}=E\boldsymbol{s}/\sqrt{(E\boldsymbol{s})^{2}}$,
and the maximum is thus given by the simple expression 
\begin{equation}
{\cal P}_{\max}=\left(E\boldsymbol{s}\right)^{2}.
\end{equation}

\subsection{Average payoff}

Following the discussion in \citep{Dakic2012}, we will assess the
average quality of the remote preparation procedure by the average
payoff $\left\langle \mathcal{P}_{\max}\right\rangle $, where the
mean value is taken over all Bloch vectors $\boldsymbol{s}$ for a
fixed direction $\boldsymbol{\beta}$ (see Fig. \ref{fig:RSP}). The
calculation of $\left\langle \mathcal{P}_{\max}\right\rangle $ can
be simplified by introducing a rotation matrix $R$ which rotates
the vector $\boldsymbol{\beta}$ onto the z axis: $\tilde{\boldsymbol{\beta}}=R\boldsymbol{\beta}=(0,0,1)$,
and an arbitrary vector $\boldsymbol{x}$ is rotated to $\tilde{\boldsymbol{x}}=R\boldsymbol{x}$.
If we further introduce the rotated correlation tensor $\tilde{E}=RER^{T}$,
we see that rotations do not change the maximal payoff: $ $
\begin{equation}
{\cal P}_{\max}=\left(E\boldsymbol{s}\right)^{2}=\left(\tilde{E}\tilde{\boldsymbol{s}}\right)^{2}.
\end{equation}
Since the vectors $\boldsymbol{s}$ and $\boldsymbol{\beta}$ are
orthogonal, the same is also true for $\tilde{\boldsymbol{s}}$ and
$\tilde{\boldsymbol{\beta}}=(0,0,1)$, and thus the normalized vector
$\tilde{\boldsymbol{s}}$ takes the form $\tilde{\boldsymbol{s}}=(\cos\phi,\sin\phi,0)$.
Using these tools we are now in position to give a closed expression
for the average payoff: 
\begin{equation}
\left\langle \mathcal{P}_{\max}\right\rangle =\frac{1}{2\pi}\int_{0}^{2\pi}d\phi\left(\tilde{E}\tilde{\boldsymbol{s}}\right)^{2}=\frac{1}{2}\mathrm{Tr}\left[E^{T}E\right]-\frac{1}{2}\left(E\boldsymbol{\beta}\right)^{2}.\label{eq:average}
\end{equation}
This expression can be proven by writing $\left(\tilde{E}\tilde{\boldsymbol{s}}\right)^{2}$
explicitly as $\left(\tilde{E}\tilde{\boldsymbol{s}}\right)^{2}=\sum_{i=1}^{3}(\tilde{E}_{i1}\cos\phi+\tilde{E}_{i2}\sin\phi)^{2}$
and evaluating the integral in Eq. (\ref{eq:average}): $\frac{1}{2\pi}\int_{0}^{2\pi}d\phi\left(\tilde{E}\tilde{\boldsymbol{s}}\right)^{2}=\frac{1}{2}\sum_{i=1}^{3}(\tilde{E}_{i1}^{2}+\tilde{E}_{i2}^{2})$.
By using the relations $\mathrm{Tr}[\tilde{E}^{T}\tilde{E}]=\sum_{i,j=1}^{3}\tilde{E}_{ij}^{2}$
and $\tilde{E}\tilde{\boldsymbol{\beta}}=(\tilde{E}_{13},\tilde{E}_{23},\tilde{E}_{33})$
the integral can further be expressed as $\frac{1}{2\pi}\int_{0}^{2\pi}d\phi\left(\tilde{E}\tilde{\boldsymbol{s}}\right)^{2}=\frac{1}{2}\mathrm{Tr}[\tilde{E}^{T}\tilde{E}]-\frac{1}{2}\left(\tilde{E}\tilde{\boldsymbol{\beta}}\right)^{2}$.
The desired equality (\ref{eq:average}) follows by noting that both
terms $\mathrm{Tr}[\tilde{E}^{T}\tilde{E}]$ and $\left(\tilde{E}\tilde{\boldsymbol{\beta}}\right)^{2}$
are invariant under rotations, i.e., $\mathrm{Tr}[\tilde{E}^{T}\tilde{E}]=\mathrm{Tr}[E^{T}E]$
and $\left(\tilde{E}\tilde{\boldsymbol{\beta}}\right)^{2}=\left(E\boldsymbol{\beta}\right)^{2}$.

\subsection{The role of quantum correlations}

Note that the matrix $E^{T}E$ has three nonnegative eigenvalues $\lambda_{1}\geq\lambda_{2}\geq\lambda_{3}$,
and the average payoff $\left\langle \mathcal{P}_{\max}\right\rangle $
satisfies the inequality 
\begin{equation}
\left\langle \mathcal{P}_{\max}\right\rangle \geq\min_{\boldsymbol{\beta}}\left\langle \mathcal{P}_{\max}\right\rangle =\frac{1}{2}(\lambda_{2}+\lambda_{3}),\label{eq:bound}
\end{equation}
where the minimum is taken over all normalized vectors $\boldsymbol{\beta}$.
This can be seen by noting that $\min_{\boldsymbol{\beta}}\left\langle \mathcal{P}_{\max}\right\rangle =\frac{1}{2}\mathrm{Tr}\left[E^{T}E\right]-\frac{1}{2}\max_{\boldsymbol{\beta}}\left(E\boldsymbol{\beta}\right)^{2}$,
and the latter maximization can be performed as $\max_{\boldsymbol{\beta}}\left(E\boldsymbol{\beta}\right)^{2}=\max_{\boldsymbol{\beta}}(\boldsymbol{\beta}E^{T}E\boldsymbol{\beta})=\lambda_{1}$
{[}see p. 176 in \citep{Horn1985}{]}. The result in Eq. (\ref{eq:bound})
is obtained by noting that $\mathrm{Tr}\left[E^{T}E\right]=\sum_{i=1}^{3}\lambda_{i}$.

According to \citep{Dakic2012}, the quantity $\min_{\boldsymbol{\beta}}\left\langle \mathcal{P}_{\max}\right\rangle $
can be regarded as the quantifier of efficiency for remote state preparation
in the worst case, i.e., for the most inconvenient choice of the direction
$\boldsymbol{\beta}$. The relation to quantum discord is established
by noting that in a large number of scenarios this expression corresponds
to the geometric measure of discord of the shared state. In general,
the geometric measure of discord is defined as \citep{Dakic2010}
\begin{equation}
D_{G}(\rho)=\min_{\sigma\in CQ}||\rho-\sigma||^{2},
\end{equation}
where the minimum is taken over all classical-quantum states $\sigma$,
and $||M||=\sqrt{\mathrm{Tr}[M^{\dagger}M]}$ is the Hilbert-Schmidt
norm of the operator $M$. For states of two qubits as given in Eq.
(\ref{eq:two qubits}) the geometric measure of discord takes the
form \citep{Dakic2010} 
\begin{equation}
D_{G}(\rho)=\frac{1}{4}(\boldsymbol{a}^{2}+\mathrm{Tr}[E^{T}E]-k_{\max}),\label{eq:D2qubits}
\end{equation}
where $\boldsymbol{a}$ is the Bloch vector of Alice's subsystem,
and $k_{\max}$ is the largest eigenvalue of the matrix $\boldsymbol{a}\boldsymbol{a}^{T}+EE^{T}$.

By the singular value decomposition of the correlation tensor $E$
it follows that the eigenvalues $\lambda_{1}\geq\lambda_{2}\geq\lambda_{3}$
of the matrix $E^{T}E$ are also eigenvalues of $EE^{T}$. Let now
$\boldsymbol{\lambda}_{1}$ be the normalized eigenvector of $EE^{T}$
corresponding to the largest eigenvalue $\lambda_{1}$ and consider
the situation where the Bloch vector $\boldsymbol{a}$ of Alice's
subsystem is parallel to $\boldsymbol{\lambda}_{1}$, i.e., $\boldsymbol{a}=\sqrt{\boldsymbol{a}^{2}}\boldsymbol{\lambda}_{1}$.
In this case the eigenvalues of the matrix $\boldsymbol{a}\boldsymbol{a}^{T}+EE^{T}$
are given as $\{\boldsymbol{a}^{2}+\lambda_{1},\lambda_{2},\lambda_{3}\}$,
and the largest eigenvalue becomes $k_{\max}=\boldsymbol{a}^{2}+\lambda_{1}$.
Inserting this result into Eq. (\ref{eq:D2qubits}) and recalling
that $\mathrm{Tr}[E^{T}E]=\sum_{i=1}^{3}\lambda_{i}$ we obtain the
expression $D_{G}(\rho)=\frac{1}{4}(\lambda_{2}+\lambda_{3})$. Together
with Eq. (\ref{eq:bound}) this result implies that for the particular
family of shared states $\rho$ where the Bloch vector $\boldsymbol{a}$
is parallel to $\boldsymbol{\lambda}_{1}$ the average payoff is bounded
below by the geometric measure of discord: 
\begin{equation}
\left\langle \mathcal{P}_{\max}\right\rangle \geq\min_{\boldsymbol{\beta}}\left\langle \mathcal{P}_{\max}\right\rangle =2D_{G}(\rho).
\end{equation}
In particular, this inequality holds for shared states $\rho$ where
the subsystem of Alice is maximally mixed. In this case the Bloch
vector of Alice is the zero vector $\boldsymbol{a}=\boldsymbol{0}$.
Another scenario satisfying the inequality is given by the states
with correlation tensor proportional to the identity matrix, i.e.,
$E_{ij}=\mu\delta_{ij}$. In this case the eigenvalues of $EE^{T}$
are all equal to $\mu^{2}$, and there is no further restriction on
the Bloch vector of Alice.

An important family of states satisfying both of these conditions
are the Werner states
\begin{equation}
\rho_{w}=p\ket{\psi^{-}}\bra{\psi^{-}}+(1-p)\frac{\openone}{4}
\end{equation}
with the singlet $\ket{\psi^{-}}=(\ket{01}-\ket{10})/\sqrt{2}$. The
state is separable for $p\leq1/3$ which can be seen by checking the
positivity of the partial transpose. As can also be seen by inspection,
the elements of the correlation tensor are given as $E_{ij}=-p\delta_{ij}$,
and the geometric measure of discord for this state becomes $D_{G}(\rho_{w})=p^{2}/2$.
In \citep{Dakic2012} these states were compared to another family
of states given by 
\begin{eqnarray}
\sigma & = & \frac{1-k}{4}\ket{\psi^{+}}\bra{\psi^{+}}+\frac{1+3k}{4}\ket{\psi^{-}}\bra{\psi^{-}}\nonumber \\
 &  & +\frac{1-2t-k}{4}\ket{00}\bra{00}+\frac{1+2t-k}{4}\ket{11}\bra{11}
\end{eqnarray}
with $\ket{\psi^{+}}=(\ket{01}+\ket{10})/\sqrt{2}$. For this family
of states it can be verified by inspection that the elements of the
correlation tensor are given as $E_{ij}=-k\delta_{ij}$, and thus
the geometric measure of discord of this state is given by $D_{G}(\sigma)=k^{2}/2$.
As was also noted in \citep{Dakic2012}, for the parameters $k=1/5$
and $t=2/5$ the state is entangled. This can be verified by calculating
the concurrence $C$ using the formula given in \citep{Wootters1998}:
$C=1/5$. 

Combining the aforementioned results, we see that the average payoff
for the separable state $\rho_{w}$ with $p=1/3$ is bounded below
as $\left\langle \mathcal{P}_{\max}\right\rangle \geq\min_{\boldsymbol{\beta}}\left\langle \mathcal{P}_{\max}\right\rangle =2D_{G}(\rho_{w})=1/9$.
In \citep{Dakic2012} this result was compared to the average payoff
achievable with the entangled state $\sigma$ for the parameters $k=1/5$
and $t=2/5$: $\left\langle \mathcal{P}_{\max}\right\rangle \geq\min_{\boldsymbol{\beta}}\left\langle \mathcal{P}_{\max}\right\rangle =2D_{G}(\sigma)=1/25$.
These results imply that for some directions $\boldsymbol{\beta}$
the separable state $\rho_{w}$ leads to a higher average payoff when
compared to the state $\sigma$, despite the fact that the latter
state is entangled.

\subsection{Discussion}

Daki\'c \emph{et al}. conclude that a shared separable state can
show a better performance for remote state preparation when compared
to entangled states \citep{Dakic2012}. In particular, if Alice and
Bob strictly follow the protocol, i.e., Alice performs von Neumann
measurements and Bob conditionally applies a $\pi$ rotation around
a given axis, there exist scenarios where shared entangled states
can be outperformed by shared states without any entanglement. As
a quantifier of the performance of the process Daki\'c \emph{et al}.
introduced a payoff function ${\cal P}$, and showed that the average
optimal payoff is bounded below by the geometric measure of discord
in a large number of scenarios. In these situations, the presence
of discord guarantees that remote state preparation can always be
achieved with nonzero average payoff. Experiment supporting these
results has also been reported \citep{Dakic2012}.

We complete the discussion by referring to the recent criticism of
this approach. On the one hand, it was shown in \citep{Giorgi2013}
that a state can lead to nonzero average payoff even if its discord
has been produced by local noise. According to \citep{Giorgi2013}
such states are unlikely to be useful in quantum information theory.
On the other hand, the restriction of the protocol to von Neumann
measurements of Alice and conditional rotations of Bob was criticized
in \citep{Tufarelli2012}. This issue was further explored in \citep{Horodecki2014},
where it was shown that by relaxing these restrictions the advantage
of separable states disappears if the standard fidelity $(1+\boldsymbol{r}\cdot\boldsymbol{s})/2$
is used as a figure of merit of the protocol. However, regardless
of this objection, it was also shown in \citep{Horodecki2014} that
in some situations separable states can still provide advantage for
remote state preparation also for the standard fidelity $(1+\boldsymbol{r}\cdot\boldsymbol{s})/2$. 

\newpage{}

\section{Entanglement distribution}

\subsection{General protocol for entanglement distribution}

\begin{figure}[H]
\noindent \begin{centering}
\includegraphics[width=0.5\columnwidth]{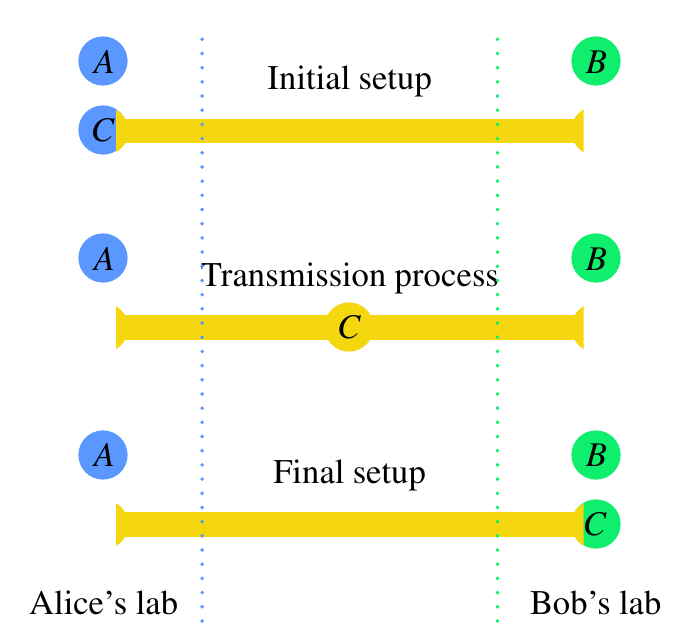}
\par\end{centering}

\caption{\label{fig:Entanglement distribution}General protocol for entanglement
distribution. A. Streltsov, H. Kampermann, and D. Bruß, \href{http://dx.doi.org/10.1103/PhysRevLett.108.250501}{Phys. Rev. Lett. \textbf{108}, 250501 (2012)}.
Copyright (2012) by the American Physical Society.}
\end{figure}
The role of quantum discord in the task of \emph{\index{Entanglement distribution}
entanglement distribution} was considered in \citep{Streltsov2012a,Chuan2012}.
The general setting is illustrated in Fig. \ref{fig:Entanglement distribution}:
Alice is initially in possession of two particles, $A$ and $C$,
while Bob is in possession of one particle $B$ (upper part of Fig.
\ref{fig:Entanglement distribution}). If Alice sends the particle
$C$ to Bob via a perfect quantum channel (middle part of Fig. \ref{fig:Entanglement distribution}),
they end up in the final setup, where Bob is in possession of both
particles $B$ and $C$, while Alice is in possession of $A$ (lower
part of Fig. \ref{fig:Entanglement distribution}).

If the total state shared by Alice and Bob is $\rho=\rho^{ABC}$,
then the initial amount of entanglement between them is given by $E^{AC|B}=E^{AC|B}(\rho)$,
while the final amount of entanglement after sending the particle
$C$ is given by $E^{A|BC}=E^{A|BC}(\rho)$. The amount of entanglement
distributed in this process is then given by the difference between
the final and the initial entanglement: $E^{A|BC}-E^{AC|B}$. In the
following, the entanglement is quantified via the relative entropy
of entanglement. For two parties $X$ and $Y$ it is defined as the
minimal relative entropy between a given state $\rho^{XY}$ and the
set of separable states ${\cal S}$: 
\begin{equation}
E^{X|Y}(\rho^{XY})=\min_{\sigma^{XY}\in{\cal S}}S(\rho^{XY}||\sigma^{XY}),\label{eq:Er}
\end{equation}
where $S(\rho||\sigma)=\mathrm{Tr}[\rho\log_{2}\rho]-\mathrm{Tr}[\rho\log_{2}\sigma]$
is the relative entropy between the states $\rho$ and $\sigma$.

\subsection{The role of quantum correlations}

The main result of \citep{Streltsov2012a,Chuan2012} is the finding
that the amount of entanglement $E^{A|BC}-E^{AC|B}$ distributed in
this protocol is limited by the amount of discord between the exchanged
particle $C$ and the rest of the system $AB$:
\begin{equation}
E^{A|BC}-E^{AC|B}\leq\Delta^{C|AB}.\label{eq:Entanglement Distribution}
\end{equation}
Here, $\Delta^{C|AB}=\Delta^{C|AB}(\rho)$ is the relative entropy
of discord, defined as the minimal relative entropy between a given
state $\rho^{XY}$ and the same state after a local von Neumann measurement:
\begin{equation}
\Delta^{X|Y}(\rho^{XY})=\min_{\left\{ \Pi_{i}^{X}\right\} }S(\rho^{XY}||\sum_{i}\Pi_{i}^{X}\rho^{XY}\Pi_{i}^{X}),\label{eq:Dr}
\end{equation}
and the minimum is taken over local von Neumann measurements $\{\Pi_{i}^{X}\}$
on the subsystem $X$.

\begin{figure}
\noindent \begin{centering}
\includegraphics[width=0.5\columnwidth]{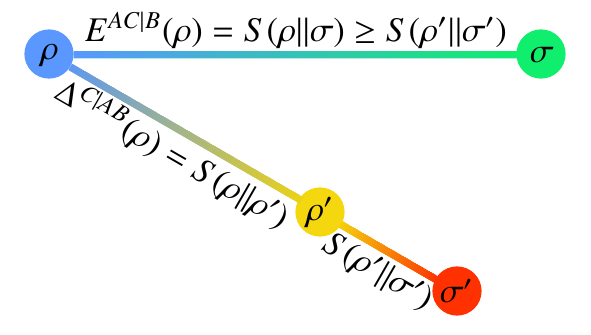}
\par\end{centering}

\caption{\label{fig:proof}Proof of the main result in Eq. (\ref{eq:Entanglement Distribution}).
A. Streltsov, H. Kampermann, and D. Bruß, \href{http://dx.doi.org/10.1103/PhysRevLett.108.250501}{Phys. Rev. Lett. \textbf{108}, 250501 (2012)}.
Copyright (2012) by the American Physical Society.}
\end{figure}
In the following we will reproduce the proof of Eq. (\ref{eq:Entanglement Distribution})
as presented in \citep{Streltsov2012a} and sketched in Fig. \ref{fig:proof}.
In particular, consider the state $\sigma$ which is separable with
respect to the bipartition $AC|B$ and at the same time the closest
separable state to $\rho$, i.e., $E^{AC|B}(\rho)=S(\rho||\sigma)$.
Moreover, define a von Neumann measurement $\{\Pi_{i}^{C}\}$ such
that the measured state $\rho'=\sum_{i}\Pi_{i}^{C}\rho\Pi_{i}^{C}$
has minimal relative entropy to $\rho$, i.e., $\Delta^{C|AB}(\rho)=S(\rho||\rho')$.
Finally, the state $\sigma'=\sum_{i}\Pi_{i}^{C}\sigma\Pi_{i}^{C}$
is defined by applying the same von Neumann measurement on the state
$\sigma$. The proof of Eq. (\ref{eq:Entanglement Distribution})
now follows by observing that the states $\rho$, $\rho'$ and $\sigma'$
lie on a straight line \citep{Streltsov2012a}: 
\begin{equation}
S(\rho||\sigma')=S(\rho||\rho')+S(\rho'||\sigma').\label{eq:proof}
\end{equation}
Before we prove this equality, we note that all quantities in this
expression are finite. In particular, $S(\rho||\rho')$ is finite
due to the definition of the state $\rho'$: $S(\rho||\rho')=\Delta^{C|AB}(\rho)$.
This also implies that the support of $\rho$ is contained in the
support of $\rho'$. Moreover, $S(\rho'||\sigma')$ is finite due
to the fact that the relative entropy does not increase under quantum
operations, and thus 
\begin{equation}
S(\rho'||\sigma')\leq S(\rho||\sigma)=E^{AC|B}(\rho).\label{eq:contractive-1}
\end{equation}
This also means that the support of $\rho'$ is contained in the support
of $\sigma'$. Combining these results we see that the support of
$\rho$ is contained in the support of $\sigma'$, and thus $S(\rho||\sigma')$
is also finite.

For proving Eq. (\ref{eq:proof}) we will first prove the following
equalities:\begin{subequations}
\begin{eqnarray}
\mathrm{Tr}[\rho\log_{2}\sigma'] & = & \mathrm{Tr}[\rho'\log_{2}\sigma'],\label{eq:proof-1}\\
\mathrm{Tr}[\rho\log_{2}\rho'] & = & \mathrm{Tr}[\rho'\log_{2}\rho'],\label{eq:proof-2}
\end{eqnarray}
\end{subequations}where all quantities are finite due to the arguments
mentioned above. Eq. (\ref{eq:proof-1}) can be proven by noting that
the state $\sigma'=\sum_{i}\Pi_{i}^{C}\sigma\Pi_{i}^{C}$ has the
form of a quantum-classical state: $\sigma'=\sum_{i}p_{i}\sigma_{i}^{AB}\otimes\Pi_{i}^{C}$
with positive probabilities $p_{i}>0$. If we further express the
states $\sigma_{i}^{AB}$ in the eigendecomposition $\sigma_{i}^{AB}=\sum_{j}\lambda_{ij}\ket{\psi_{ij}}\bra{\psi_{ij}}^{AB}$
with positive eigenvalues $\lambda_{ij}>0$ and eigenstates $\ket{\psi_{ij}^{AB}}$,
we obtain the following: \begin{subequations} 
\begin{eqnarray}
\mathrm{Tr}[\rho'\log_{2}\sigma'] & = & \mathrm{Tr}\left[\left(\sum_{k}\Pi_{k}^{C}\rho\Pi_{k}^{C}\right)\left(\sum_{ij}\log_{2}(p_{i}\lambda_{ij})\ket{\psi_{ij}}\bra{\psi_{ij}}^{AB}\otimes\Pi_{i}^{C}\right)\right]\\
 & = & \sum_{ijk}\log_{2}(p_{i}\lambda_{ij})\mathrm{Tr}\left[\Pi_{k}^{C}\rho\Pi_{k}^{C}\ket{\psi_{ij}}\bra{\psi_{ij}}^{AB}\otimes\Pi_{i}^{C}\right]\label{eq:linear}\\
 & = & \sum_{ijk}\log_{2}(p_{i}\lambda_{ij})\mathrm{Tr}\left[\rho\ket{\psi_{ij}}\bra{\psi_{ij}}^{AB}\otimes\Pi_{k}^{C}\Pi_{i}^{C}\Pi_{k}^{C}\right]\label{eq:cyclic}\\
 & = & \sum_{ij}\log_{2}(p_{i}\lambda_{ij})\mathrm{Tr}\left[\rho\ket{\psi_{ij}}\bra{\psi_{ij}}^{AB}\otimes\Pi_{i}^{C}\right]\label{eq:orthogonal}\\
 & = & \mathrm{Tr}\left[\rho\sum_{ij}\log_{2}(p_{i}\lambda_{ij})\ket{\psi_{ij}}\bra{\psi_{ij}}^{AB}\otimes\Pi_{i}^{C}\right]=\mathrm{Tr}[\rho\log_{2}\sigma'].\label{eq:linear-1}
\end{eqnarray}
\end{subequations} In Eq. (\ref{eq:linear}) we used the linearity
of the trace, and in Eq. (\ref{eq:cyclic}) its cyclic invariance.
In Eq. (\ref{eq:orthogonal}) we used the orthogonality of projectors,
i.e., $\Pi_{k}^{C}\Pi_{i}^{C}\Pi_{k}^{C}=\delta_{ki}\Pi_{i}^{C}$.
By applying the linearity of the trace once again in Eq. (\ref{eq:linear-1})
we arrive at the desired result: $\mathrm{Tr}[\rho\log_{2}\sigma']=\mathrm{Tr}[\rho'\log_{2}\sigma']$.
Using the same arguments Eq. (\ref{eq:proof-2}) is also seen to be
correct.

The proof of Eq. (\ref{eq:proof}) now follows by applying these results
to the sum $S(\rho||\rho')+S(\rho'||\sigma')$: 
\begin{eqnarray}
S(\rho||\rho')+S(\rho'||\sigma') & = & \mathrm{Tr}[\rho\log_{2}\rho]-\mathrm{Tr}[\rho\log_{2}\rho']+\mathrm{Tr}[\rho'\log_{2}\rho']-\mathrm{Tr}[\rho'\log_{2}\sigma']\nonumber \\
 & \overset{\mathrm{Eq.\,(\ref{eq:proof-2})}}{=} & \mathrm{Tr}[\rho\log_{2}\rho]-\mathrm{Tr}[\rho'\log_{2}\rho']+\mathrm{Tr}[\rho'\log_{2}\rho']-\mathrm{Tr}[\rho'\log_{2}\sigma']\nonumber \\
 & = & \mathrm{Tr}[\rho\log_{2}\rho]-\mathrm{Tr}[\rho'\log_{2}\sigma']\nonumber \\
 & \overset{\mathrm{Eq.\,(\ref{eq:proof-1})}}{=} & \mathrm{Tr}[\rho\log_{2}\rho]-\mathrm{Tr}[\rho\log_{2}\sigma']=S(\rho||\sigma').
\end{eqnarray}

We now turn to the proof of the main result in Eq. (\ref{eq:Entanglement Distribution}).
Starting from Eq. (\ref{eq:proof}) and recalling that the state $\rho'$
was defined such that $S(\rho||\rho')=\Delta^{C|AB}(\rho)$ we obtain
the following equality: $S(\rho||\sigma')=\Delta^{C|AB}(\rho)+S(\rho'||\sigma')$.
In the next step we make use of Eq. (\ref{eq:contractive-1}) arriving
at the following result:
\begin{equation}
S(\rho||\sigma')\leq\Delta^{C|AB}(\rho)+E^{AC|B}(\rho).\label{eq:proof-3}
\end{equation}
In the final step, recall that the state $\sigma$ is separable with
respect to the bipartition $AC|B$, and thus can be written as $\sigma=\sum_{j}q_{j}\sigma_{j}^{AC}\otimes\sigma_{j}^{B}$.
Using this expression, we can write the state $\sigma'=\sum_{i}\Pi_{i}^{C}\sigma\Pi_{i}^{C}$
as 
\begin{equation}
\sigma'=\sum_{ij}p_{ij}q_{j}\sigma_{ij}^{A}\otimes\sigma_{j}^{B}\otimes\Pi_{i}^{C}
\end{equation}
with $\sum_{i}\Pi_{i}^{C}\sigma_{j}^{AC}\Pi_{i}^{C}=\sum_{i}p_{ij}\sigma_{ij}^{A}\otimes\Pi_{i}^{C}$.
From this result we see that the state $\sigma'$ is fully separable,
and thus the relative entropy between $\rho$ and $\sigma'$ is an
upper bound on the relative entropy of entanglement $E^{A|BC}$, i.e.,
$E^{A|BC}(\rho)\leq S(\rho||\sigma')$. Inserting this inequality
into Eq. (\ref{eq:proof-3}) we arrive at the desired result: $E^{A|BC}(\rho)\leq\Delta^{C|AB}(\rho)+E^{AC|B}(\rho)$.
This completes the proof of Eq. (\ref{eq:Entanglement Distribution}).

As was further pointed out in \citep{Streltsov2012a}, the inequality
(\ref{eq:Entanglement Distribution}) also holds in a more general
case, where the relative entropy $S$ in both equations (\ref{eq:Er})
and (\ref{eq:Dr}) is replaced by a general distance $D$ which has
the following properties:
\begin{itemize}
\item $D$ does not increase under quantum operations, i.e., 
\begin{equation}
D(\Lambda[\rho],\Lambda[\sigma])\leq D(\rho,\sigma)
\end{equation}
 for any quantum operation $\Lambda$ and any pair of states $\rho$
and $\sigma$,
\item $D$ satisfies the triangle inequality, i.e., 
\begin{equation}
D(\rho,\sigma)\leq D(\rho,\tau)+D(\tau,\sigma)
\end{equation}
 for any states $\rho$, $\sigma$, and $\tau$.
\end{itemize}
By virtue of the triangle inequality, Eq. (\ref{eq:proof}) changes
to the inequality 
\begin{equation}
D(\rho,\sigma')\leq D(\rho,\rho')+D(\rho',\sigma').
\end{equation}
Starting from this result, Eq. (\ref{eq:Entanglement Distribution})
can be proven following the same reasoning as for the relative entropy.
Important examples for distances having these properties are the trace
distance $D_{t}(\rho_{1},\rho_{2})=\frac{1}{2}\mathrm{Tr}|\rho_{1}-\rho_{2}|$
with the trace norm of an operator $M$ defined as $\mathrm{Tr}|M|=\mathrm{Tr}\sqrt{M^{\dagger}M}$
and the Bures distance $D_{B}(\rho_{1},\rho_{2})=2(1-\sqrt{F(\rho_{1},\rho_{2})})$
with the fidelity $F(\rho_{1},\rho_{2})=\left(\mathrm{Tr}\sqrt{\sqrt{\rho_{1}}\rho_{2}\sqrt{\rho_{1}}}\right)^{2}$.

\subsection{Discussion}

The result in Eq. (\ref{eq:Entanglement Distribution}) reveals a
fundamental relation between the amount of entanglement in different
splits of a quantum system. This can be seen by permuting the parties
$A$ and $B$ in Eq. (\ref{eq:Entanglement Distribution}), which
delivers the following inequality \citep{Chuan2012}: 
\begin{equation}
\left|E^{A|BC}-E^{AC|B}\right|\leq\Delta^{C|AB}.
\end{equation}
This inequality provides a strong link between $E^{A|BC}$ and $E^{AC|B}$.
In particular, for zero quantum discord $\Delta^{C|AB}=0$ this result
immediately implies that these quantities are equal: $E^{A|BC}=E^{AC|B}$. 

Finally, we notice that for successful entanglement distribution the
exchanged particle does not need to be entangled with the rest of
the system. In particular, there exist states $\rho=\rho^{ABC}$ which
exhibit no entanglement between the exchanged particle $C$ and the
rest of the system $AB$, i.e, 
\begin{equation}
\rho=\sum_{i}p_{i}\cdot\rho_{i}^{AB}\otimes\rho_{i}^{C},
\end{equation}
while at the same time the final amount of entanglement $E^{A|BC}$
is strictly larger than the initial amount of entanglement $E^{AC|B}$:
$E^{A|BC}(\rho)>E^{AC|B}(\rho)$. The possibility of such \emph{\index{Entanglement distribution!with separable states}
entanglement distribution with separable states} was first pointed
out by Cubitt \emph{et al}. in \citep{Cubitt2003}, who proved that
this phenomenon is possible with vanishing initial entanglement $E^{AC|B}(\rho)=0$.
In the last years these results were further extended to different
classes of quantum states \citep{Mista2008,Chuan2012,Kay2012,Park2012},
and a classical counterpart for this quantum phenomenon was also presented
\citep{Bae2009}. The limits for this effect were further explored
in \citep{Streltsov2013}, where it was shown that entanglement distribution
with separable states requires states with rank at least three if
the amount of entanglement is quantified via the logarithmic negativity.
Very recently, three independent experiments have also shown that
entanglement distribution with separable states is indeed possible
with current technology \citep{Fedrizzi2013,Vollmer2013,Peuntinger2013,Silberhorn2013}. 

\newpage{}

\section{Transmission of correlations}

\subsection{Classical transmission of correlations}

\begin{figure}[h]
\noindent \begin{centering}
\includegraphics[width=1\textwidth]{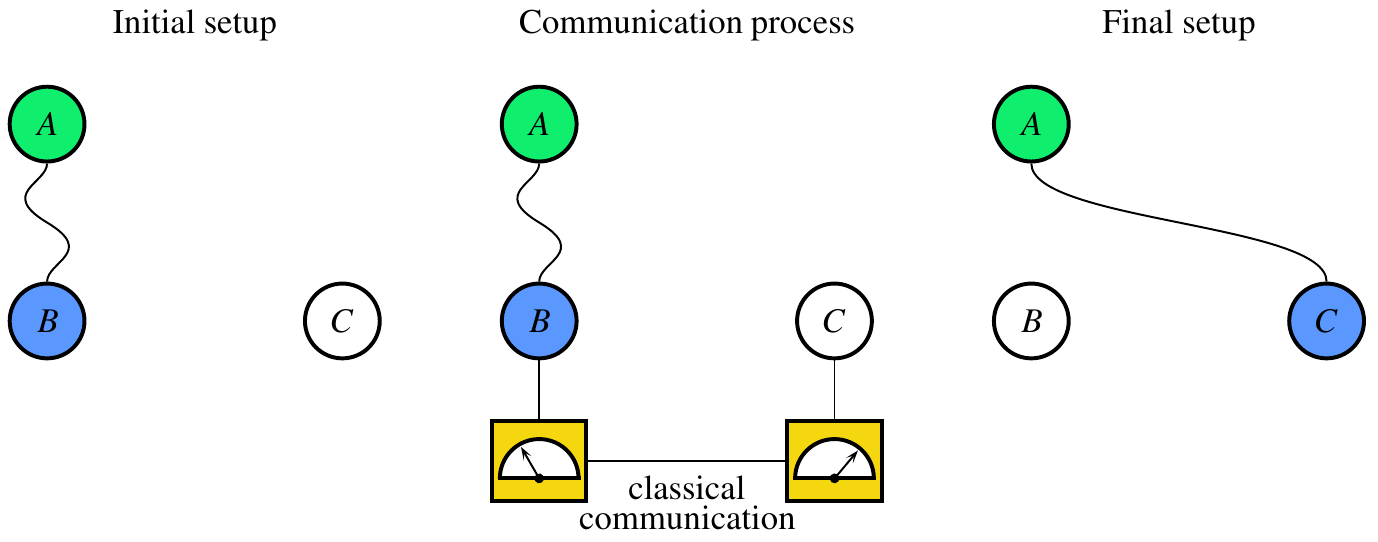}
\par\end{centering}

\caption{\label{fig:classical-transmission}Classical transmission of correlations.}
\end{figure}
The role of quantum discord for the \emph{\index{Transmission of correlations!classical}
}transmission of correlations was studied in \citep{Streltsov2013a}.
The setup is illustrated in Fig. \ref{fig:classical-transmission}:
initially Alice and Bob share a joint quantum state $\rho^{AB}$.
The third party Charlie is initially uncorrelated with Alice and Bob,
i.e., the total initial state is given by 
\begin{equation}
\rho^{ABC}=\rho^{AB}\otimes\rho^{C},
\end{equation}
see left part of Fig. \ref{fig:classical-transmission}. In the task
of \emph{classical transmission} Bob aims to transfer his state to
Charlie by the means of local operations and classical communication
(LOCC), see middle part of Fig. \ref{fig:classical-transmission}.
After this process the final state takes the form 
\begin{equation}
\rho_{f}^{ABC}=\Lambda_{B\leftrightarrow C}[\rho^{ABC}],
\end{equation}
where $\Lambda_{B\leftrightarrow C}$ denotes an LOCC operation between
Bob and Charlie. In the ideal case the final state $\rho_{f}^{AC}$
shared by Alice and Charlie is equal to the initial state $\rho^{AB}$
shared by Alice and Bob (right part of Fig. \ref{fig:classical-transmission}):
\begin{equation}
\rho_{f}^{AC}=\rho^{AB}.\label{eq:transmission}
\end{equation}

As was pointed out in \citep{Streltsov2013a}, such an ideal process
is not possible in general. In particular, due to the fact that entanglement
cannot be created by LOCC, the final state $\rho_{f}^{AC}$ is always
separable. This implies that Eq. (\ref{eq:transmission}) is never
fulfilled if Alice and Bob share an entangled initial state $\rho^{AB}$.
As was further shown in \citep{Streltsov2013a}, the ideal process
is possible if and only if Alice and Bob share a quantum-classical
state: 
\begin{equation}
\rho^{AB}=\sum_{i}p_{i}\rho_{i}^{A}\otimes\ket{i}\bra{i}^{B}.
\end{equation}
This was shown by introducing a figure of merit $I^{c}$ which quantifies
the maximal mutual information between Alice and Charlie achievable
in this procedure. The formal definition of $I^{c}$ can be given
as follows: 
\begin{equation}
I^{c}(\rho^{AB})=\lim_{d_{C}\rightarrow\infty}\sup_{\Lambda_{B\leftrightarrow C}}I(\rho_{f}^{AC}),\label{eq:Ic}
\end{equation}
where the supremum is taken over all LOCC operations $\Lambda_{B\leftrightarrow C}$
between Bob and Charlie, $I$ is the mutual information, and $d_{C}$
is the dimension of Charlie's system.

\subsection{The role of quantum correlations}

As was shown in \citep{Streltsov2013a}, the figure of merit for the
classical transmission $I^{c}$ introduced in Eq. (\ref{eq:Ic}) is
closely related to the amount of discord in the initial state $\rho^{AB}$.
The latter is defined as 
\begin{equation}
D^{B|A}(\rho^{AB})=I(\rho^{AB})-\sup_{\{M_{i}^{B}\}}J(\rho^{AB})_{\{M_{i}^{B}\}},\label{eq:discord-1}
\end{equation}
where $J(\rho^{AB})_{\{M_{i}^{B}\}}$ is given as 
\begin{equation}
J(\rho^{AB})_{\{M_{i}^{B}\}}=S(\rho^{A})-\sum_{i}p_{i}S(\rho_{i}^{A}).\label{eq:J-1}
\end{equation}
Here, $\{M_{i}^{B}\}$ is a positive operator-valued measure (POVM)
on Bob's system $B$, $p_{i}=\mathrm{Tr}[M_{i}^{B}\rho^{AB}]$ is
the probability for the outcome $i$, and $\rho_{i}^{A}=\mathrm{Tr}_{B}[M_{i}^{B}\rho^{AB}]/p_{i}$
is the state of Alice after the outcome $i$ has been obtained. $I^{c}$
is related to the discord $D^{B|A}$ as follows:
\begin{equation}
I^{c}(\rho^{AB})=I(\rho^{AB})-D^{B|A}(\rho^{AB}).\label{eq:IcMain}
\end{equation}
This equality was proven in \citep{Streltsov2013a}, and we will present
an alternative proof in the following. In particular we will prove
the inequalities 
\begin{eqnarray}
I^{c}(\rho^{AB}) & \leq & I(\rho^{AB})-D^{B|A}(\rho^{AB}),\label{eq:<}\\
I^{c}(\rho^{AB}) & \geq & I(\rho^{AB})-D^{B|A}(\rho^{AB}),\label{eq:>}
\end{eqnarray}
which taken together imply Eq. (\ref{eq:IcMain}).

For proving Eq. (\ref{eq:<}) we consider the structure of the final
state $\rho_{f}^{ABC}=\Lambda_{B\leftrightarrow C}[\rho^{ABC}]$ using
the fact that any LOCC operation $\Lambda_{B\leftrightarrow C}$ can
be written as a separable operation%
\footnote{The inverse is not true in general, i.e., a separable operation does
not necessarily correspond to LOCC \citep{Horodecki2009}.%
}
\begin{equation}
\rho_{f}^{ABC}=\Lambda_{B\leftrightarrow C}[\rho^{ABC}]=\sum_{i=1}^{m}B_{i}\otimes C_{i}\rho^{ABC}B_{i}^{\dagger}\otimes C_{i}^{\dagger}
\end{equation}
with a finite number of terms $m$ and Kraus operators $B_{i}\otimes C_{i}$
satisfying $\sum_{i=1}^{m}B_{i}^{\dagger}B_{i}\otimes C_{i}^{\dagger}C_{i}=\openone^{B}\otimes\openone^{C}$
\citep{Horodecki2009}. In the next step recall that the initial state
$\rho^{ABC}$ has the form $\rho^{ABC}=\rho^{AB}\otimes\rho^{C}$.
Moreover, $I^{c}$ does not depend on the choice of the state $\rho^{C}$,
and thus we choose $\rho^{C}=\openone^{C}/d_{C}$. With this in mind,
the final state $\rho_{f}^{ABC}$ can also be written as
\begin{equation}
\rho_{f}^{ABC}=\frac{1}{d_{C}}\sum_{i=1}^{m}B_{i}\rho^{AB}B_{i}^{\dagger}\otimes C_{i}C_{i}^{\dagger}.
\end{equation}
Now we define positive numbers $q_{i}=\mathrm{Tr}[C_{i}C_{i}^{\dagger}]>0$
and quantum states $\sigma_{i}^{C}=C_{i}C_{i}^{\dagger}/q_{i}$. The
expression for the final state $\rho_{f}^{ABC}$ further reduces to
\begin{equation}
\rho_{f}^{ABC}=\sum_{i=1}^{m}E_{i}^{B}\rho^{AB}\left(E_{i}^{B}\right)^{\dagger}\otimes\sigma_{i}^{C}.\label{eq:rhof-1}
\end{equation}
Here, $E_{i}^{B}$ are Kraus operators on the subsystem $B$ defined
as $E_{i}^{B}=\sqrt{\frac{q_{i}}{d_{C}}}B_{i}$. The fact that $E_{i}^{B}$
are indeed Kraus operators, i.e., satisfy $\sum_{i=1}^{m}\left(E_{i}^{B}\right)^{\dagger}E_{i}^{B}=\openone^{B}$,
can be verified by inspection: 
\begin{eqnarray}
\sum_{i=1}^{m}\left(E_{i}^{B}\right)^{\dagger}E_{i}^{B} & = & \sum_{i=1}^{m}\frac{q_{i}}{d_{C}}B_{i}^{\dagger}B_{i}=\frac{1}{d_{C}}\sum_{i=1}^{m}\mathrm{Tr}[C_{i}C_{i}^{\dagger}]\cdot B_{i}^{\dagger}B_{i}\\
 & = & \frac{1}{d_{C}}\mathrm{Tr}_{C}\left[\sum_{i=1}^{m}B_{i}^{\dagger}B_{i}\otimes C_{i}^{\dagger}C_{i}\right]=\frac{1}{d_{C}}\mathrm{Tr}_{C}\left[\openone^{B}\otimes\openone^{C}\right]=\openone^{B}.\nonumber 
\end{eqnarray}

Starting from the result in Eq. (\ref{eq:rhof-1}) the final state
shared by Alice and Charlie takes the form 
\begin{equation}
\rho_{f}^{AC}=\sum_{i=1}^{m}\mathrm{Tr}_{B}\left[M_{i}^{B}\rho^{AB}\right]\otimes\sigma_{i}^{C},\label{eq:rhof-2}
\end{equation}
where $M_{i}^{B}$ are POVM elements on the subsystem $B$ defined
as $M_{i}^{B}=\left(E_{i}^{B}\right)^{\dagger}E_{i}^{B}$. We will
now show that for any such state the mutual information is bounded
above as follows: 
\begin{equation}
I(\rho_{f}^{AC})\leq I(\rho^{AB})-D^{B|A}(\rho^{AB}).\label{eq:<-1}
\end{equation}
 Since the figure of merit $I^{c}$ was defined as the supremum of
the mutual information between Alice and Charlie over all LOCC protocols
in the limit $d_{C}\rightarrow\infty$, this result will imply the
inequality (\ref{eq:<}). To prove this statement we introduce the
state 
\begin{equation}
\tau^{A\tilde{C}}=\sum_{i=1}^{m}\mathrm{Tr}_{B}\left[M_{i}^{B}\rho^{AB}\right]\otimes\ket{i}\bra{i}^{\tilde{C}}\label{eq:tau}
\end{equation}
with a new system $\tilde{C}$ having dimension $d_{\tilde{C}}=\max\{d_{C},m\}$.
Note that the latter state can be transformed into the state $\Lambda_{\tilde{C}}[\tau^{A\tilde{C}}]=\sum_{i=1}^{m}\mathrm{Tr}_{B}\left[M_{i}^{B}\rho^{AB}\right]\otimes\sigma_{i}^{\tilde{C}}$
by a local operation%
\footnote{The local operation that achieves this task is a measure-and-prepare
map with Kraus operators $K_{ab}^{\tilde{C}}=\sqrt{\sigma_{b}^{\tilde{C}}}\ket{a}\bra{b}^{\tilde{C}}$.%
} $\Lambda_{\tilde{C}}$, where the states $\sigma_{i}^{\tilde{C}}$
are the same as $\sigma_{i}^{C}$ in Eq. (\ref{eq:rhof-2}). Since
the mutual information does not increase under local operations, it
follows that the state $\tau^{A\tilde{C}}$ has at least the same
mutual information as $\rho_{f}^{AC}$: $ $
\begin{equation}
I(\tau^{A\tilde{C}})\geq I(\rho_{f}^{AC}).
\end{equation}
Finally, it is straightforward to verify that the mutual information
of $\tau^{A\tilde{C}}$ can be written as 
\begin{equation}
I(\tau^{A\tilde{C}})=J(\rho^{AB})_{\{M_{i}^{B}\}}
\end{equation}
 with $J(\rho^{AB})_{\{M_{i}^{B}\}}$ defined in Eq. (\ref{eq:J-1}).
Combining these results we arrive at the inequality 
\begin{equation}
I(\rho_{f}^{AC})\leq I(\tau^{A\tilde{C}})\leq\sup_{\{M_{i}^{B}\}}J(\rho^{AB})_{\{M_{i}^{B}\}}=I(\rho^{AB})-D^{B|A}(\rho^{AB}),
\end{equation}
where the last equality follows from the definition of discord in
Eq. (\ref{eq:discord-1}). This completes the proof of Eq. (\ref{eq:<-1})
and the inequality (\ref{eq:<}).

We will now complete the proof of Eq. (\ref{eq:IcMain}) by proving
the inequality (\ref{eq:>}). This can be done by considering a specific
LOCC protocol where Bob performs a measurement with $d_{C}$ Kraus
operators $E_{i}^{B}$ on his subsystem $B$. The outcome of the measurement
is sent to Charlie who stores it in his system $C$ of dimension $d_{C}$.
After performing this protocol, the final state $\rho_{f}^{AC}$ shared
by Alice and Charlie takes the form
\begin{equation}
\rho_{f}^{AC}=\sum_{i=1}^{d_{C}}\mathrm{Tr}_{B}\left[M_{i}^{B}\rho^{AB}\right]\otimes\ket{i}\bra{i}^{C}
\end{equation}
with POVM elements $M_{i}^{B}=\left(E_{i}^{B}\right)^{\dagger}E_{i}^{B}$.
Since we consider a specific LOCC protocol, $I_{c}$ cannot be smaller
than the mutual information for any state obtained in this way, and
thus 
\begin{equation}
I^{c}(\rho^{AB})\geq\lim_{d_{C}\rightarrow\infty}\sup_{\{M_{i}^{B}\}}I(\rho_{f}^{AC}).
\end{equation}
The inequality (\ref{eq:>}) follows by noting that the state $\rho_{f}^{AC}$
has the same form as $\tau^{A\tilde{C}}$ in Eq. (\ref{eq:tau}),
and thus by applying the same arguments as for $\tau^{A\tilde{C}}$
we see that the mutual information of $\rho_{f}^{AC}$ can be written
as $I(\rho_{f}^{AC})=J(\rho^{AB})_{\{M_{i}^{B}\}}$. Together with
the definition of discord in Eq. (\ref{eq:discord-1}) this completes
the proof of Eqs. (\ref{eq:>}) and (\ref{eq:IcMain}).

\subsection{Quantum transmission of correlations}

\begin{figure}
\noindent \begin{centering}
\includegraphics[width=1\textwidth]{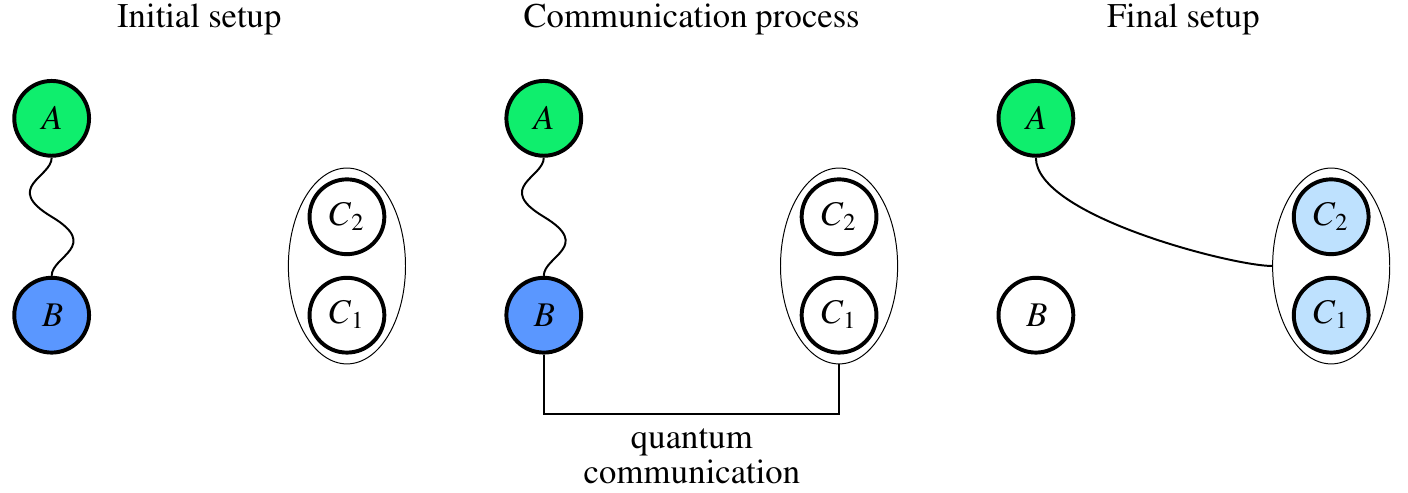}
\par\end{centering}

\caption{\label{fig:quantum-transmission}Quantum transmission of correlations.}
\end{figure}

The task of \index{Transmission of correlations!quantum} \emph{quantum
transmission} was considered in \citep{Streltsov2013a} and independently
in \citep{Brandao2013}. The setup is illustrated in Fig. \ref{fig:quantum-transmission}.
Similar to the scenario for the classical transmission, Alice and
Bob share a joint initial state $\rho^{AB}$ (left part of Fig. \ref{fig:quantum-transmission}).
The system of Charlie now consists of two subsystems $C_{1}$ and
$C_{2}$, initially uncorrelated with Alice and Bob, and the total
initial state is thus given by 
\begin{equation}
\rho^{ABC}=\rho^{AB}\otimes\rho^{C_{1}C_{2}}.
\end{equation}
Moreover, Bob and Charlie have access to a general quantum communication
channel $\Lambda_{BC}$ (middle part of Fig. \ref{fig:quantum-transmission}),
and the final state after the application of the channel takes the
form 
\begin{equation}
\rho_{f}^{ABC}=\Lambda_{BC}[\rho^{ABC}],
\end{equation}
see right part of Fig. \ref{fig:quantum-transmission}. 

The aim of this process is to achieve maximal mutual information between
the system of Alice and each of Charlie's subsystems $C_{1}$ and
$C_{2}$ on average. Following \citep{Streltsov2013a} we denote the
corresponding figure of merit by $I_{2}^{q}$, where the superscript
$q$ tells us that quantum communication is considered, and the index
$2$ gives the number of Charlie's subsystems. The formal definition
of $I_{2}^{q}$ can be given as follows: 
\begin{equation}
I_{2}^{q}(\rho^{AB})=\lim_{d\rightarrow\infty}\sup_{\Lambda_{BC}}\frac{I(\rho_{f}^{AC_{1}})+I(\rho_{f}^{AC_{2}})}{2},
\end{equation}
where both subsystems of Charlie have the same dimension $d=d_{C_{1}}=d_{C_{2}}$.
It is straightforward to generalize this quantity to $n$ subsystems
of Charlie: 
\begin{equation}
I_{n}^{q}(\rho^{AB})=\lim_{d\rightarrow\infty}\sup_{\Lambda_{BC}}\frac{\sum_{i=1}^{n}I(\rho_{f}^{AC_{i}})}{n},\label{eq:Iq}
\end{equation}
where $d=d_{C_{1}}=d_{C_{2}}=\ldots=d_{C_{n}}$ is the dimension of
each of Charlie's subsystems.

\subsection{Equivalence of quantum and classical transmission for pure states}

In the scenario where Alice and Bob share a pure initial state $\rho^{AB}=\ket{\psi}\bra{\psi}^{AB}$
quantum and classical transmission are equivalent \citep{Streltsov2013a}:
\begin{equation}
I_{n}^{q}(\ket{\psi}\bra{\psi}^{AB})=I^{c}(\ket{\psi}\bra{\psi}^{AB})\label{eq:equivalence}
\end{equation}
for any number of Charlie's subsystems $n\geq2$. We will present
the proof for this statement following the arguments of \citep{Streltsov2013a}.
First, it is important to note that $I_{n}^{q}$ cannot be smaller
than $I^{c}$: 
\begin{equation}
I_{n}^{q}(\rho^{AB})\geq I^{c}(\rho^{AB}),\label{eq:Iq>In}
\end{equation}
which follows from the fact that quantum communication is more general
than classical communication. In the following we will show that for
pure states $\ket{\psi}^{AB}$ and $n\geq2$ the inverse inequality
also holds:
\begin{equation}
I_{n}^{q}(\ket{\psi}\bra{\psi}^{AB})\leq I^{c}(\ket{\psi}\bra{\psi}^{AB}).\label{eq:In<Iq}
\end{equation}
This result together with Eq. (\ref{eq:Iq>In}) will complete the
proof of the desired equality (\ref{eq:equivalence}). 

In the first step we will show that the sum $\sum_{i=1}^{n}I(\rho_{f}^{AC_{i}})$
is in general bounded above as follows: 
\begin{equation}
\sum_{i=1}^{n}I(\rho_{f}^{AC_{i}})\leq nS(\rho_{f}^{A}).\label{eq:<-2}
\end{equation}
This inequality can be proven by using the fact that any tripartite
state $\rho^{XYZ}$ satisfies the inequality 
\begin{equation}
I(\rho^{XY})+I(\rho^{XZ})\leq2S(\rho^{X}),
\end{equation}
which can be seen by rewriting it as $S(\rho^{Y})+S(\rho^{Z})\leq S(\rho^{XY})+S(\rho^{XZ})$
and noting that the latter inequality is equivalent to the strong
subadditivity of the von Neumann entropy {[}see p. 521 in \citep{Nielsen2000}{]}.
If we apply this inequality to the state $\rho_{f}^{AC_{k}C_{l}}$
with $k\neq l$, we arrive at the following inequality:
\begin{equation}
I(\rho_{f}^{AC_{k}})+I(\rho_{f}^{AC_{l}})\leq2S(\rho_{f}^{A}).\label{eq:subadditivity-1}
\end{equation}
Starting from this result we will now prove Eq. (\ref{eq:<-2}) for
even $n\geq2$, i.e., $n=2m$. In this case the sum can be bounded
as 
\begin{equation}
\sum_{i=1}^{n}I(\rho_{f}^{AC_{i}})=\sum_{j=1}^{m}\left\{ I(\rho_{f}^{AC_{2j-1}})+I(\rho_{f}^{AC_{2j}})\right\} \leq2mS(\rho_{f}^{A})=nS(\rho_{f}^{A}),\label{eq:subadditivity-2}
\end{equation}
where we used Eq. (\ref{eq:subadditivity-1}) to obtain $I(\rho_{f}^{AC_{2j-1}})+I(\rho_{f}^{AC_{2j}})\leq2S(\rho_{f}^{A})$.
For odd $n\geq3$ we can write 
\begin{eqnarray}
2\sum_{i=1}^{n}I(\rho_{f}^{AC_{i}}) & = & \sum_{k=1}^{n}I(\rho_{f}^{AC_{k}})+\sum_{l=1}^{n}I(\rho_{f}^{AC_{l}})\\
 &  & =I(\rho_{f}^{AC_{1}})+\sum_{k=2}^{n}I(\rho_{f}^{AC_{k}})+\sum_{l=1}^{n-1}I(\rho_{f}^{AC_{l}})+I(\rho_{f}^{AC_{n}})\nonumber \\
 &  & =\sum_{k=2}^{n}I(\rho_{f}^{AC_{k}})+\sum_{l=1}^{n-1}I(\rho_{f}^{AC_{l}})+I(\rho_{f}^{AC_{1}})+I(\rho_{f}^{AC_{n}}).\nonumber 
\end{eqnarray}
Note that the sums $\sum_{k=2}^{n}I(\rho_{f}^{AC_{k}})$ and $\sum_{l=1}^{n-1}I(\rho_{f}^{AC_{l}})$
each have $n-1$ terms, which is an even number. Thus, we can use
the same arguments as in Eq. (\ref{eq:subadditivity-2}) to see that
both of them are bounded from above by $(n-1)S(\rho_{f}^{A})$. Finally,
due to Eq. (\ref{eq:subadditivity-1}) the sum $I(\rho_{f}^{AC_{1}})+I(\rho_{f}^{AC_{n}})$
is bounded from above by $2S(\rho_{f}^{A})$. Combining these results
we see that
\begin{equation}
2\sum_{i=1}^{n}I(\rho_{f}^{AC_{i}})\leq2nS(\rho_{f}^{A}),
\end{equation}
which completes the proof of the desired inequality (\ref{eq:<-2})
for any $n\geq2$.

Recalling that the state of Alice never changes in the process, i.e.,
$\rho_{f}^{A}=\rho^{A}$, Eq. (\ref{eq:<-2}) implies that the average
mutual information $\frac{1}{n}\sum_{i=1}^{n}I(\rho_{f}^{AC_{i}})$
never exceeds the entropy of $\rho^{A}$: 
\begin{equation}
\frac{1}{n}\sum_{i=1}^{n}I(\rho_{f}^{AC_{i}})\leq S(\rho^{A}).
\end{equation}
Due to the definition of $I_{n}^{q}$ in Eq. (\ref{eq:Iq}) this result
immediately implies that $S(\rho^{A})$ is also an upper bound for
$I_{n}^{q}$:
\begin{equation}
I_{n}^{q}(\rho^{AB})\leq S(\rho^{A}).
\end{equation}
In the final step, note that for pure states $\rho^{AB}=\ket{\psi}\bra{\psi}^{AB}$
the quantity $I^{c}$ coincides with the entropy of $\rho^{A}$: 
\begin{equation}
I^{c}(\ket{\psi}\bra{\psi}^{AB})=S(\rho^{A}).
\end{equation}
This follows from the relation between $I^{c}$ and quantum discord
provided in Eq. (\ref{eq:IcMain}) by noting that for pure states
the mutual information and the discord are given as $I(\ket{\psi}\bra{\psi}^{AB})=2S(\rho^{A})$
and $D^{B|A}(\ket{\psi}\bra{\psi}^{AB})=S(\rho^{A})$. This completes
the proof of Eq. (\ref{eq:In<Iq}) and the desired equality (\ref{eq:equivalence})
immediately follows.

\subsection{Discussion}

As was pointed out in \citep{Streltsov2013a}, ideal classical transmission
as illustrated in Fig. \ref{fig:classical-transmission} is possible
if and only if the initial state $\rho^{AB}$ is quantum-classical:
\begin{equation}
\rho^{AB}=\sum_{i}p_{i}\rho_{i}^{A}\otimes\ket{i}\bra{i}^{B}.
\end{equation}
For any other state the discord $D^{B|A}(\rho^{AB})$ is nonzero,
and the process of classical transmission unavoidably leads to a loss
of information, i.e., the mutual information between Alice and Charlie
is never larger than the difference $I(\rho^{AB})-D^{B|A}(\rho^{AB})$.
The amount of quantum discord $D^{B|A}(\rho^{AB})$ thus quantifies
the loss of information in the task of classical transmission.

It was further shown in \citep{Streltsov2013a} that the equivalence
between quantum and classical transmission stated in Eq. (\ref{eq:equivalence})
only holds for pure initial state. In particular, there exist mixed
states $\rho^{AB}$ for which quantum transmission leads to a better
performance when compared to the classical transmission: $I_{2}^{q}(\rho^{AB})>I^{c}(\rho^{AB})$.
In this context, an important result was obtained recently by Brandão
\emph{et al}., who showed that $I_{n}^{q}$ and $I^{c}$ coincide
in the asymptotic limit \citep{Brandao2013}: 
\begin{equation}
\lim_{n\rightarrow\infty}I_{n}^{q}(\rho^{AB})=I^{c}(\rho^{AB}).
\end{equation}
Thus, quantum and classical transmission are equivalent for any initial
state $\rho^{AB}$ if the number of Charlie's subsystems goes to infinity.

\chapter{Outlook}

In this work, we discussed the role of quantum correlations beyond
entanglement in three fundamental tasks in quantum information theory:
remote state preparation \citep{Dakic2012}, entanglement distribution
\citep{Streltsov2012a,Chuan2012}, and transmission of correlations
\citep{Streltsov2013a,Brandao2013}. Although these tasks clearly
demonstrate the relevance of quantum discord and general quantum correlations
in quantum information theory, they cannot cover the whole range of
applications of quantum correlations beyond entanglement that have
been presented recently. In the following, we will give an outlook
on some of the developments in this direction.

The role of quantum discord in \emph{\index{Quantum!metrology} quantum
metrology} was first investigated by Modi \emph{et al}. \citep{Modi2011},
and important contributions in this direction were made recently by
Girolami \emph{et al}. in \citep{Girolami2013,Girolami2014}. In the
scenario considered in \citep{Girolami2014}, Alice and Bob share
a bipartite state $\rho^{AB}$ undergoing a local unitary evolution
$U_{A}=e^{-i\varphi H_{A}}$ on the subsystem of Alice with a nondegenerate
Hamiltonian $H_{A}$. The final state $U_{A}\rho^{AB}U_{A}^{\dagger}$
is then used to estimate the unknown parameter $\varphi$. As was
shown in \citep{Girolami2014}, the parameter $\varphi$ can always
be estimated with nonzero precision whenever the state $\rho^{AB}$
is not classical-quantum, i.e., not of the form $\rho^{AB}=\sum_{i}p_{i}\ket{i}\bra{i}^{A}\otimes\rho_{i}^{B}$.
The authors of \citep{Girolami2014} investigate this phenomenon by
introducing a new quantifier of quantum correlations which they call
\emph{interferometric power.} They show that the interferometric power
is able to capture the worst-case precision of the procedure, and
conclude that the presence of discord in a quantum state guarantees
its usefulness for quantum metrology. Experiment supporting these
theoretical results has also been reported in \citep{Girolami2014}.

A great amount of attention was also attracted by the relation between
entanglement and discord in the quantum measurement process \citep{Streltsov2011,Piani2011,Gharibian2011}.
In particular, it was shown in \citep{Streltsov2011,Piani2011} that
for performing a von Neumann measurement on one part of a composite
quantum state $\rho^{AB}$, the creation of entanglement between the
system and the measurement apparatus is unavoidable whenever the state
has nonzero quantum discord. Recently, experimental demonstration
of this effect has also been reported \citep{Adesso2014}. These results
support the role of quantum discord and general quantum correlations
for studying entanglement on the one hand, and for understanding phenomena
which cannot be explained solely by the presence of entanglement on
the other hand. In this context, useful results can be expected from
the investigation of quantum discord in the framework of coherence,
recently introduced by Baumgratz \emph{et al}. \citep{Baumgratz2013}.
The main aim of this research direction would be the unification of
all three concepts: entanglement, quantum correlations beyond entanglement,
and coherence. This research may further lead to the discovery of
new tasks in quantum information theory which are not based on entanglement,
and which require new types of quantum correlations to capture their
performance.

\bibliographystyle{naturemag}
\bibliography{literature}

\end{document}